\documentclass[chicago]{emulateapj}
\usepackage[section]{placeins}
\usepackage{amsmath}
\usepackage{graphicx}
\usepackage[dvips]{color}
\slugcomment{Draft v5-- Jan152016}
\shortauthors{}

\begin{document}

\title{Detection of Linearly Polarized 6.9 mm Continuum Emission from the Class 0 Young Stellar Object NGC1333 IRAS4A}

\author{Hauyu Baobab Liu\altaffilmark{1}} \author{Shih-Ping Lai \altaffilmark{2}} \author{Yasuhiro Hasegawa \altaffilmark{3,4,5}} \author{Naomi Hirano\altaffilmark{1}} \author{Ramprasad Rao \altaffilmark{1}} \author{I-Hsiu Li\altaffilmark{1}} \author{Misato Fukagawa\altaffilmark{3}} \author{Josep M. Girart\altaffilmark{6,7}} \author{Carlos Carrasco-Gonz\'{a}lez\altaffilmark{8}} \author{Luis F. Rodr\'{i}guez\altaffilmark{8}}

\affil{$^{1}$Academia Sinica Institute of Astronomy and Astrophysics, P.O. Box 23-141, Taipei, 106 Taiwan}\email{hyliu@asiaa.sinica.edu.tw}

\affil{$^{2}$Institute of Astronomy and Department of Physics, National Tsing Hua University, Hsinchu, Taiwan}

\affil{$^{3}$National Astronomical Observatory of Japan, Mitaka, Tokyo 181-8588, Japan}

\affil{$^{4}$EACOA Fellow}

\affil{$^{5}$Jet Propulsion Laboratory, California Institute of Technology, Pasadena, CA 91109, USA}

%\affil{$^{6}$Graduate School of Science, Osaka University, 1-1 Machikaneyama, Toyonaka, Osaka 560-0043}

\affil{$^{6}$Institut de Ci\`{e}ncies de l'Espai, (CSIC-IEEC),Campus UAB, Carrer de Can Magrans, S/N, 08193 Cerdanyola del Vall{\'e}s, Catalonia, Spain}

\affil{$^{7}$Harvard-Smithsonian Center for Astrophysics, 60 Garden Street, Cambridge, MA 02138, USA}

\affil{$^{8}$Instituto de Radioastronom{\'i}a y Astrof{\'i}sica, UNAM, A.P. 3-72, Xangari, Morelia, 58089, Mexico}

\begin{abstract}
We report new Karl G. Jansky Very Large Array (JVLA),  $0\farcs5$ angular resolution observations of linearly polarized continuum emission at 6.9 mm, towards the Class 0 young stellar object (YSO) NGC1333 IRAS4A. 
This target source is a collapsing dense molecular core, which was resolved at short wavelengths to have hourglass shaped B-field configuration.
We compare these 6.9 mm observations with previous polarization Submillimeter Array (SMA) observations at 0.88 mm, which have comparable angular resolution ($\sim$$0\farcs7$).
We found that at the same resolution, the observed polarization position angles at 6.9 mm are slightly deviated from those observed at 0.88 mm.
Due to the lower optical depth of the emission at 6.9 mm, and the potential effect of dust grain growth, the new JVLA observations are likely probing B-field alignments in regions interior to those sampled by the previous polarization observations at higher frequencies. 
% We tentatively interpret the observed polarization position angles (rotated by 90$^{\circ}$) as tracing the direction of the B-field pinched inside the geometrically flattened accretion flow. 
Our understanding can be improved by more sensitive observations, and observations for the more extended spatial scales.
\end{abstract}


% this work, measured at the peak of polarized intensity (0''.49x0''.4)
% before residual removal: B-field: ~84 deg
% after residual removal: B-field: ~67 deg

% cite Hirota et al. (2008) for the distance

%============================================================
\clearpage
\section{Introduction }
\label{chap_introduction}
Multi-wavelength observations of Stokes I dust thermal emission from young stellar objects (YSOs) have provided invaluable information for coagulation of dust grains (e.g. Kwon et al. 2009; Ricci et al. 2010; Guilloteau et al. 2011; Chiang et al. 2012; P{\'e}rez et al. 2012, 2015; Ricci et al. 2012; see also Testi et al. 2014 for a review).
Polarized dust emission, on the other hand, involves alignment of dust grains that are non-spherical in shape. 
Although the mechanism of dust alignment is still inconclusive, the standard theories suggest that the long axes of anisotropic dust grains are aligned perpendicular to the magnetic (B-) field lines (Hildebrand 1988; Lazarian et al. 1997). 
Since dust grains close to the spherically symmetric shape should emit less polarized light, observations of polarized dust emission may simultaneously provide information about the morphology of growing dust grains as well as the B-field orientation in circumstellar environments, which are two of the most intriguing aspects in this research field (for more discussion see Hughes et al. 2009, 2013).

Previous interferometric 3 mm, 1 mm, and 0.88 mm observations of YSOs have successfully detected polarized dust thermal emission on $\ge$300 AU scales, which is considered to trace the pinched poloidal B-field  in the collapsing low-mass circumstellar envelopes (e.g. IRAS 16293-2422, $d\sim$120 pc: Akeson 
\& Carlstrom 1997, Rao et al. 2009; NGC1333 IRAS4A, $d\sim$235 pc: Akeson \& Carlstrom 1997, Girart et al. 2006, Gon{\c c}alves et 
al. 2008, Frau et al. 2011; TADPOL survey: Hull et al. 2014).
Furthermore, the recent sub-arcsecond resolution millimeter and submillimeter observations have inferred the toroidal B-field components in the circumstellar disks embedded in the two Class 0 YSOs IRAS 16293-2422B (Rao et al. 2014), in the Class 0 YSO L1527 (Segura-Cox et al. 2015), and in the T Tauri object HL Tau (Stephens et al. 2014). 
% Furthermore, the recent $\sim$$0\farcs5$ resolution SMA observations have succeeded in observing the 0.88 mm polarized dust emission towards the circumstellar disks embedded in the two Class 0 YSOs IRAS 16293-2422B (Rao et al. 2014).
% The observed polarized emission from smaller scales infers that the B-field in the disks is very likely to be predominantly toroidal, which can arise from dragging of B-fields by differential rotation.   
The toroidal B-field components can arise from dragging of B-fields by differential rotation.
% The fiducial analytical calculations based on the assumptions of perfectly toroidal B-field and radiative-torque grain alignment have predicted that the polarization fraction at wavelengths larger than 100 $\mu$m is typically 2\%-3\% in the T Tauri disks (Cho \& Lazarian 2007), which is consistent with previous observations (Rao et al. 2014).
In addition, the suggestions of toroidal B-fields in disks were further strengthened by the discoveries of the Keplerian rotating disks in Class 0 YSOs (e.g. Choi et al. 2010; Tobin et al. 2012; Murillo et al. 2013; Ohashi et al. 2014).

\begin{table*}
\caption{Records of Q-band Polarization Observations on J0713+4349 and 3C84$^{1, 2}$}
\label{tab-polcal}
\hspace{3cm}
\vspace{-3cm}
\begin{tabular}{l l c c c}\hline\hline
Source & Date	&	Stokes I	Flux &	Polarized Flux		&	Polarization Percentage		\\
		&		&	(Jy)				&  (Jy)					&  (\%)								\\\hline
J0713+4349	& 2010 Oct. 29	&	0.3801	&	0.0003	&	0.09 \\
J0713+4349	& 2010 Dec. 15	&  0.4331	&	0.0007	&	0.16	\\
J0713+4349	& 2011 Mar. 04	&  0.3491	&	0.0094	&	2.69  \\
J0713+4349	& 2011 Nov. 03	&  0.3783	&	0.0002	&	0.06  \\
J0713+4349	& 2012 Feb. 03	&  0.5542 	&	0.0001	&	0.02  \\
J0713+4349	& 2012 Feb. 25	&  0.5967	&	0.0001	&	0.02  \\
J0713+4349	& 2012 Apr. 05	&  0.3939 	&	0.0013 	&	0.32  \\
J0713+4349	& 2012 Jun. 03	&  0.5100	&	0.0002	&	0.04  \\\hline
3C84				& 2012 Feb. 03	&	40.5439	&	0.0010	&	0.0025	\\
3C84				& 2012 Mar. 05	&	33.0175 	&	0.0465	&	0.14		\\
3C84				& 2012 Apr. 08	&	27.1929 	&	0.0013	&	0.0047	\\
3C84				& 2012 May. 27	&	23.4269 	&	0.0034	&	0.01		\\\hline
\end{tabular}
\vspace{3cm}
\caption*{\footnotesize{
\hspace{0cm}
\begin{tabular}{ p{17cm} }
$^{1}$\,Data are quoted from the Master EVLA POLCAL Database ({\tt http://www.aoc.nrao.edu/$\sim$smyers/evlapolcal/polcal\_master.html}). \\
$^{2}$\,3C84 at this moment approaches $\sim$1\% polarization at 43 GHz, according to the Guide to Observing with the VLA ({\tt https://science.nrao.edu/facilities/vla/docs/manuals/obsguide/modes/pol}). \\ 
\end{tabular}
\vspace{0.5cm}
}}
\end{table*}

The present project intends to further explore the capability of the upgraded National Radio Astronomy Observatory (NRAO)\footnote{The National Radio Astronomy Observatory is a facility of the National Science Foundation operated under cooperative agreement by Associated Universities, Inc.} Karl G. Jansky Very Large Array (JVLA) in detecting dust polarization at wavelengths longer than 6 mm.
For YSOs embedded in dense gas, long wavelength observations are expected to preferentially trace large dust grains, which form more efficiently in the center, higher density regions.
% Therefore, it can probe magnetic field configuration on the small scales (i.e. in the disks), without being significantly confused by the extended envelope.
Therefore, these observations can probe the magnetic field configuration on the smaller scales (i.e., the disks), without being significantly confused by the extended envelope.
This will be of paramount importance in understanding how Keplerian disks form in the magnetically threaded accreting envelope (e.g. Zhao et al. 2011; Kataoka et al. 2012;  Hennebelle \& Ciardi 2009; Krasnopolsky et al. 2012; Krumholz et al. 2013; Li et al. 2013; Seifried et al. 2013), and may permit the investigation of B-field configurations in deeply embedded massive star-forming disks (e.g. Liu et al. 2013).
In addition, dust emission from YSOs is in general optically thinner at longer wavelengths, which are less affected by the polarization/de-polarization due to scattering.
Moreover, from the observed polarization intensity/percentage, we are able to gauge the feasibility of carrying out follow-up observations using the JVLA (or the Band 1 observations of the Atacama Large Millimeter Array), with the highest possible angular resolution ($\sim$$0\farcs05$; $\sim$6.3 AU assuming $d\sim$130 pc) in the future. 
For regions with a non-uniform B-field morphology, the high angular resolution imaging enabled by the unprecedented sensitivity is essential to avoid canceling of the polarized emission due to spatial smearing. 

We selected to observe the extensively studied Class 0 YSO, NGC1333 IRAS4A (e.g. Girart et al. 1999, 2006).
We compare our JVLA observations with the existing SMA\footnote{The Submillimeter Array is a joint project between the Smithsonian Astrophysical Observatory and the Academia Sinica Institute of Astronomy and Astrophysics, and is funded by the Smithsonian Institution and the Academia Sinica (Ho, Moran, \& Lo 2004).} (Ho, Moran, \& Lo 2004; Marrone 2006) polarization observations at 0.88 mm (Girart et al. 2006; Frau et al. 2011; Lai et al. in prep.)
Our JVLA observations and the SMA data are described in Section \ref{chap_obs}.
The results are presented in Section \ref{chap_result}.
A brief discussion is provided in Section \ref{chap_discussion}.
Our conclusions and remarks are given in Section \ref{chap_summary}.

%============================================================

% JVLA image after uv_10_455 limit: 0.73x0.61; -84.8
% SMA image after uv_10_455 limit: 0.74x0.61; 78.4

\section{Observations and Data Reduction} 
\label{chap_obs}
\subsection{The JVLA Data}
\label{sub_jvladata}
\subsubsection{Observations and Basic Calibrations}
\label{subsub:jvlacal}
The JVLA observations of linearly polarized emission at 6.2-7.5 mm (Q band) can be simply calibrated, and are not subjected to the spurious circular polarization due to the beam squint effect (e.g. EVLA memo 113). 
This results from the fact that the Q and U parameters that define linear polarization are relatively unaffected by beam squint when using circular feeds (Hamaker et al. 1996; Sault et al. 1996).
We have performed the JVLA Q band observations towards NGC1333 IRAS4A in C array configurations on 2014 October 13 (project code: 14B-053).
We took full RR, RL, LR, and LL correlator products. 
These observations had an overall duration of 120 minutes, with 57 minutes of integration on target source. 
After initial data flagging, 26 antennas were available.
The projected baseline lengths covered by these observations are in the range of 36-3140 meters ($\sim$5-460 $k\lambda$), which yielded $\sim$$0\farcs5$ angular resolution, and  a maximum detectable angular scale (i.e. the recovered flux $\sim$1/$e$ of the original flux) of $\sim$21$''$ ($\sim$5170 AU; Wilner \& Welch 1994).
We used the 3 bit sampler, and configured the backend to have an 8 GHz bandwidth coverage by 64 consecutive spectral windows, which were centered on the sky frequency of 44 GHz.
The expected root-mean-square (RMS) noise level for our Stokes I continuum image incorporating data in the 8 GHz bandwidth, is $\sim$15 $\mu$Jy\,beam$^{-1}$.
The pointing center for our target source is on R.A.=03$^{\mbox{\scriptsize{h}}}$29$^{\mbox{\scriptsize{m}}}$10.550$^{\mbox{\scriptsize{s}}}$ (J2000), Decl.=+31$^{\circ}$13$'$31$''$.0 (J2000).
The complex gain calibrator J0336+3218 was observed for 54 seconds every 162 seconds to calibrate the atmospheric and instrumental gain phase and amplitude  fluctuation.
We observed the bright but weakly polarized ($\sim$1\%) quasar 3C84 for 171 seconds for passband and polarization leakage (i.e. D-term) calibrations. 
We integrated on the standard calibrator 3C147 for 174 seconds for referencing the absolute flux scaling and the polarization position angle.
We further integrated on an essentially non-polarized faint source J0713+4349 for 231 seconds, for measuring the residual polarization leakage after implementing the D-term solution derived from 3C84 (more in Section \ref{subsub:jvlapol}).

\begin{figure*}
\vspace{-1cm}
\begin{tabular}{p{8.5cm} p{8.5cm} }
\includegraphics[width=10.25cm]{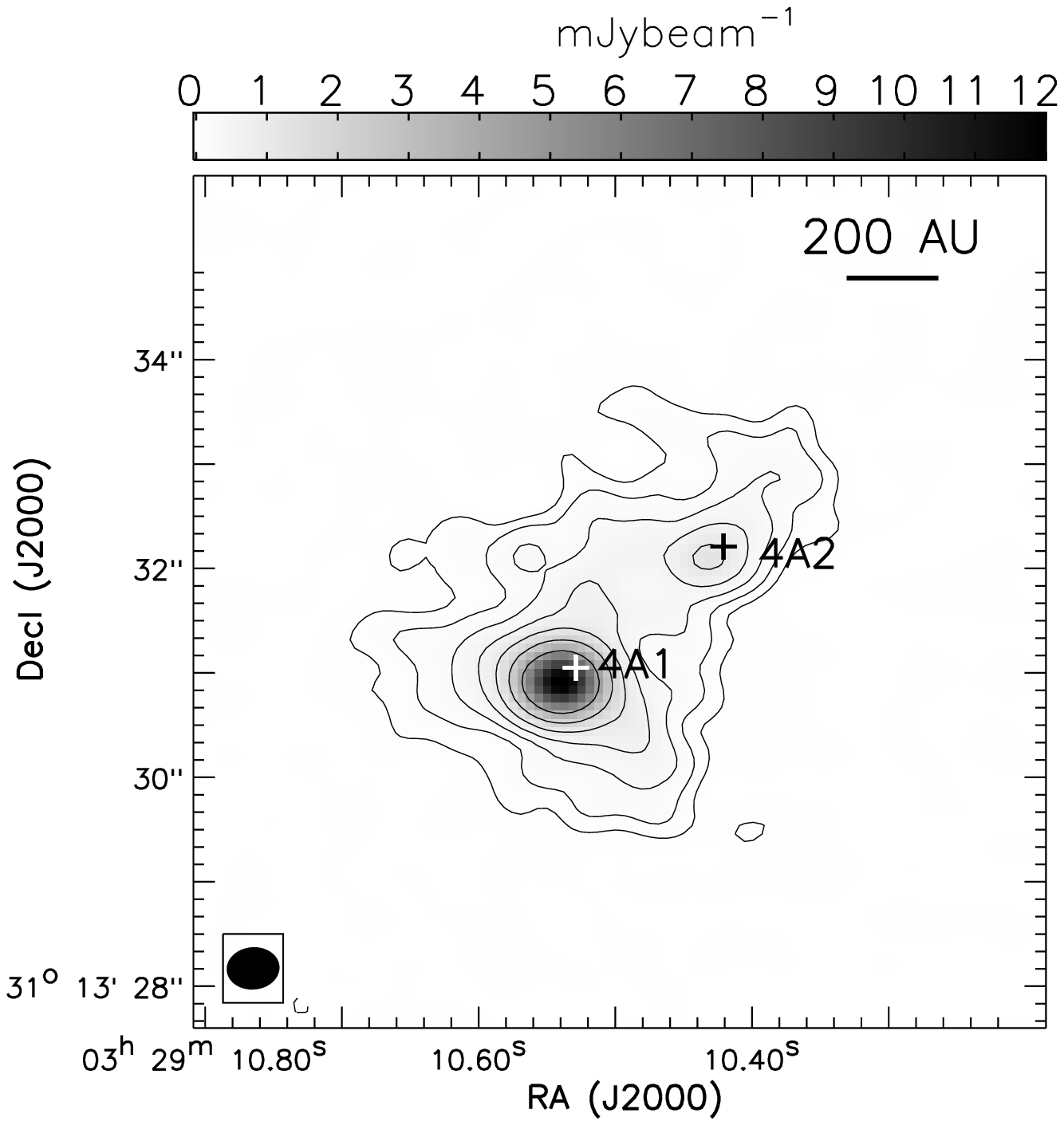} & \includegraphics[width=10.25cm]{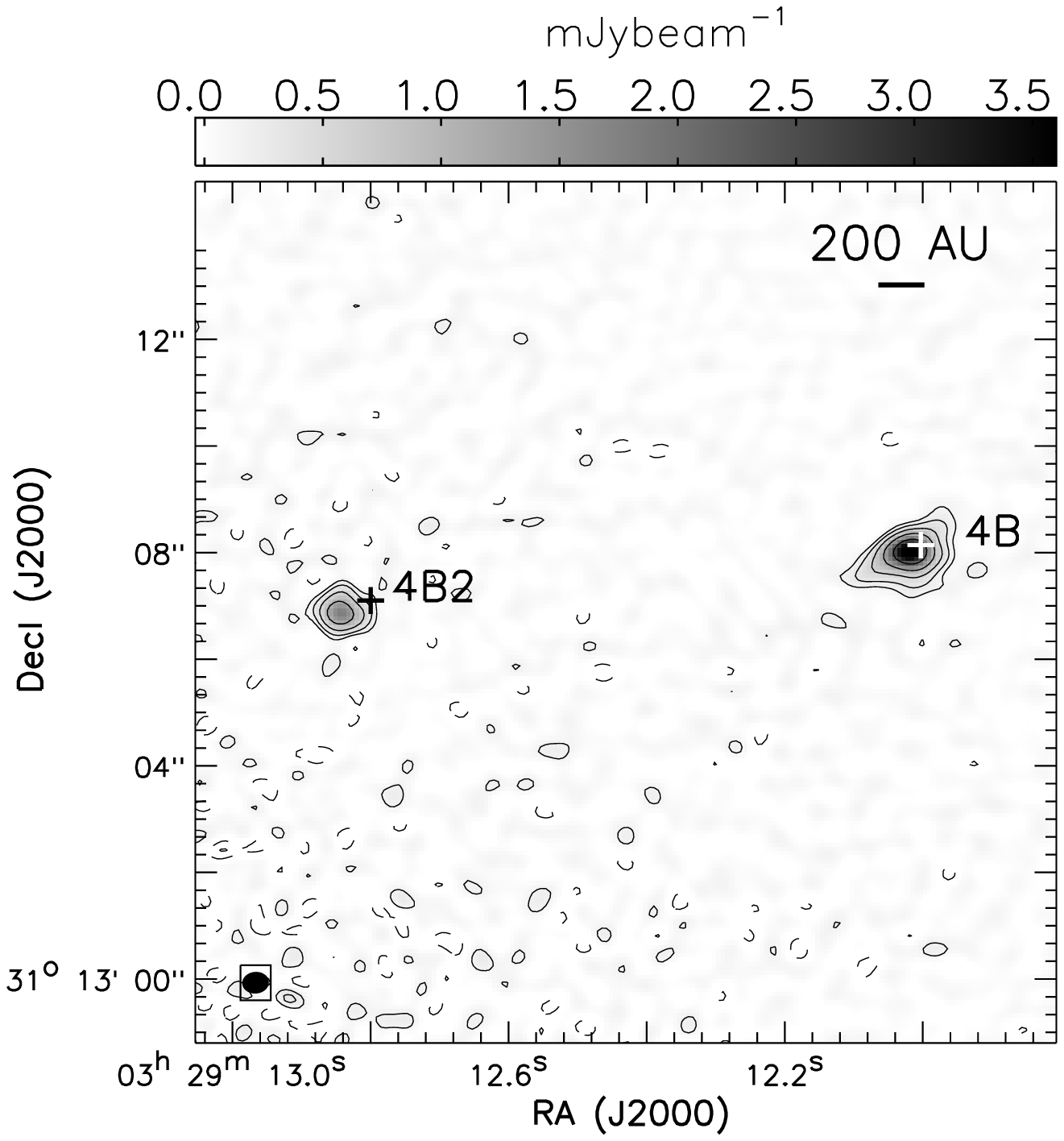} \\
\end{tabular}
\caption{\footnotesize{
The JVLA 6.9 mm Stokes I continuum images of NGC1333 IRAS 4A and 4B. The synthesized beam is $\theta_{\mbox{\scriptsize{maj}}}$$\times$$\theta_{\mbox{\scriptsize{min}}}$=$0\farcs49$$\times$$0\farcs40$ (P.A.=-85$^{\circ}$). Contours are 69 $\mu$Jy\,beam$^{-1}$ (3$\sigma$; $\sim$0.188 K) $\times$ [-2, -1, 1, 2, 4, 8, 16, 32, 64], and 141.0 $\mu$Jy\,beam$^{-1}$ (3$\sigma$; $\sim$0.386 K) $\times$ [-2, -1, 1, 2, 4, 8, 16] for the left and right panels, respectively. The higher noise level for the right panel is because the covered region is closer to the edge of the primary beam. The coordinates of the "+" symbols were quoted from Reipurth et al. (2002), and Hull et al. (2014).
}}
\vspace{0.4cm}
\label{fig:jvlastokesI}
\end{figure*}

The past records of the polarization observations on our polarization leakage calibrators are provided in Table \ref{tab-polcal}.
We manually followed the standard data calibration strategy using the Common Astronomy Software Applications (CASA; McMullin et al. 2007) package release 4.2.2 (release 30986).
After implementing the antenna position corrections, weather information, gain-elevation curve and opacity model, we bootstrapped delay fitting and passband calibrations, and then performed complex gain calibration, cross-hand delay fitting, polarization leakage calibration (using 3C84), and polarization position angle referencing.
We applied the absolute flux reference to our complex gain solutions, and then applied all derived solution tables to the target source and J0713+4349.
Finally, we performed 3 iterations of gain phase self-calibrations for our target source, to remove the residual phase offsets, in particular, phase jumps, and then performed one additional iteration of simultaneous gain amplitude and phase self-calibration.

We generated images using the CASA task \texttt{clean}. 
The image size is 1600 pixels in each dimension, and the pixel size is $0\farcs08$.
The achieved synthesized beam in the Briggs Robust=0 weighted image is $\theta_{\mbox{\scriptsize{maj}}}$$\times$$\theta_{\mbox{\scriptsize{min}}}$=$0\farcs49$$\times$$0\farcs40$ (P.A.=-84.6$^{\circ}$).
The observed peak flux density of Stokes I continuum emission is 12.7 mJy\,beam$^{-1}$ ($\sim$41 K).
The measured RMS noise level in the Stokes I, Q, and U images are $\sim$23 $\mu$Jy\,beam$^{-1}$.
The measured noise level is higher than the theoretical noise level, probably as a result of data flagging during the calibration and self-calibration processes.
% The higher measured noise level than the theoretical noise level is probably due to data flagging during the calibration and self-calibration processes. 

\subsubsection{Removing Residual Polarization Leakages}
\label{subsub:jvlapol}
Polarization leakage solutions can either be derived based on a short integration of an unpolarized source, or by tracking a polarized calibrator with wide parallactic angle coverages (Sault et al. 1996).
Considering the realistic weather condition of the JVLA site, high frequency ($>$30 GHz) observations often have to be carried out in short-duration ($\le$2 hrs) scheduling blocks.
Therefore, the former calibration strategy may be more practical than the latter one, and will permit a more flexible scheduling of observations. 

Observing bright calibration sources for deriving polarization leakage solutions is preferable for short-duration scheduling blocks, for the sake of minimizing overhead ratio.
We note that although the broad bandwidth coverages of the JVLA observations can provide good continuum Stokes I sensitivity, for accurately calibrating the polarization observations, it is required to achieve high signal-to-noise (S/N) ratios in small chunks of frequencies, to gauge the bandwidth decorrelation and depolarization effects (see Hamaker et al. 1996). 
However, most of the bright calibration sources, usually quasars, are known to be polarized. 
As a compromise, we first derived the polarization leakage solutions  for individual spectral channels based on the observations of the weakly polarized  ($\sim$1\%) bright quasar 3C84, and then measured the residual polarization leakage from the full Stokes images of the faint, essentially unpolarized source J0713+4349.

\begin{figure*}
\hspace{-0.4cm}
%\vspace{-9cm}
\begin{tabular}{ p{4cm} p{4cm} p{4cm} p{4cm} }
\includegraphics[width=6cm]{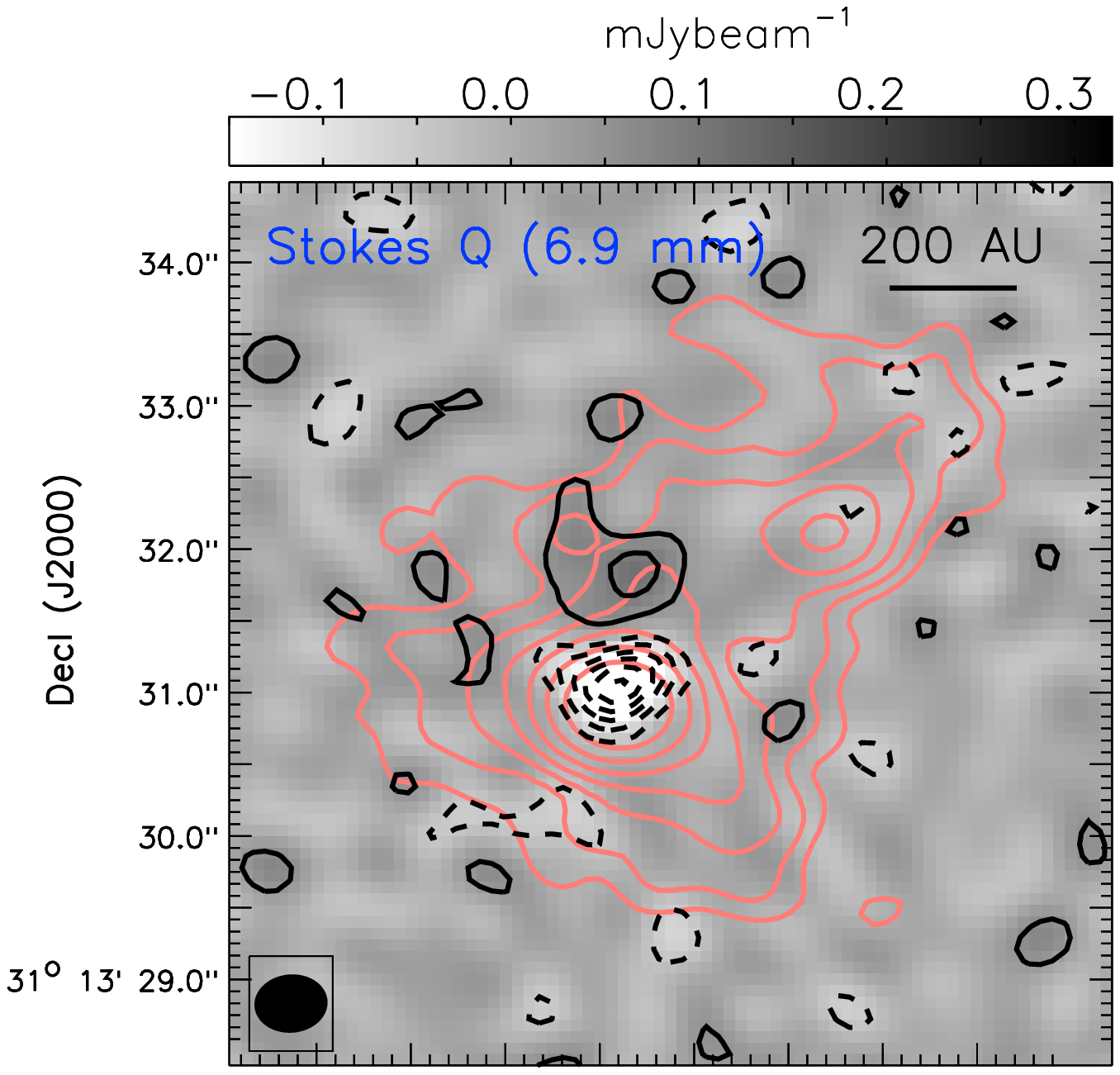} & \includegraphics[width=6cm]{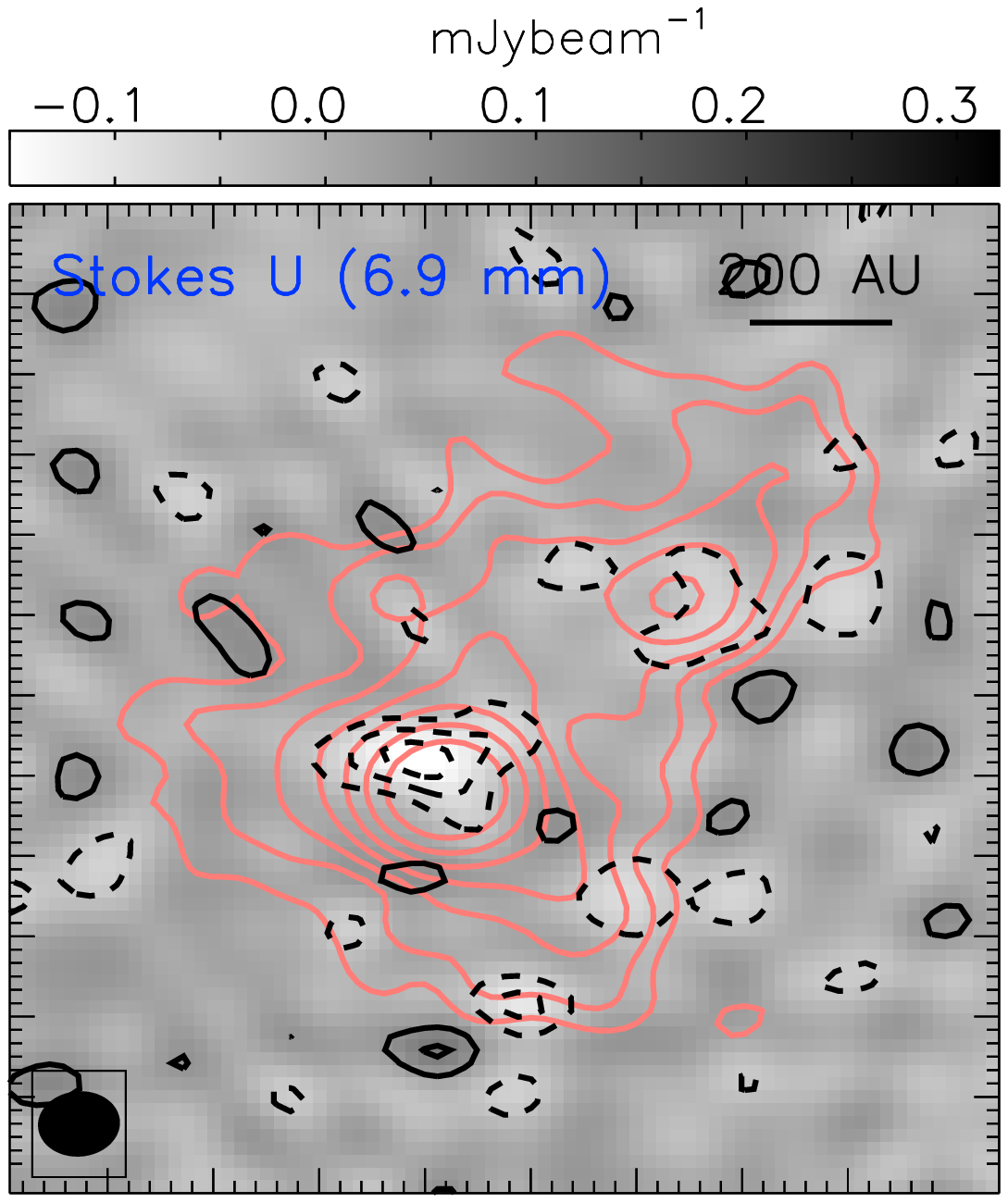} & \includegraphics[width=6cm]{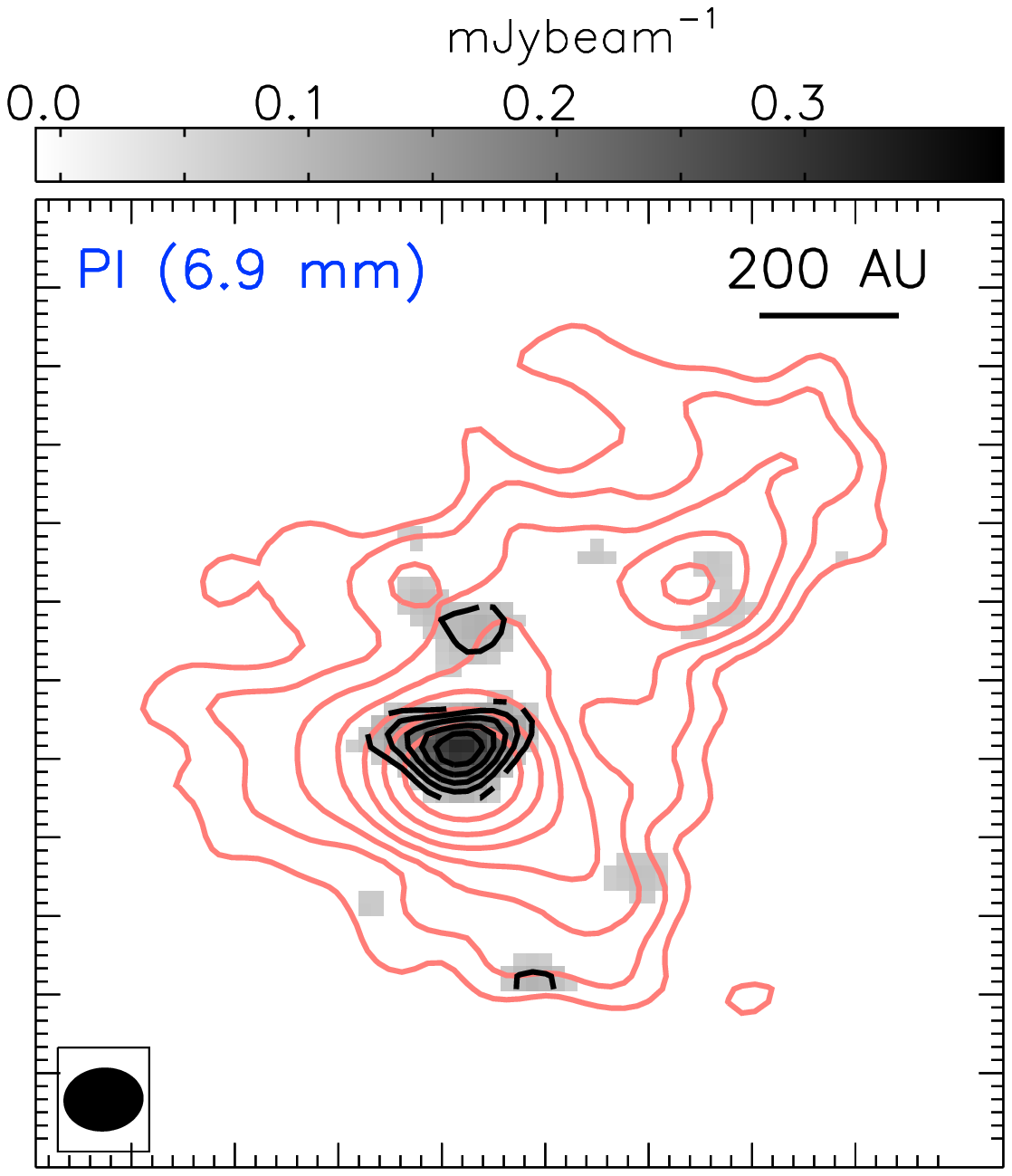}  & \includegraphics[width=6cm]{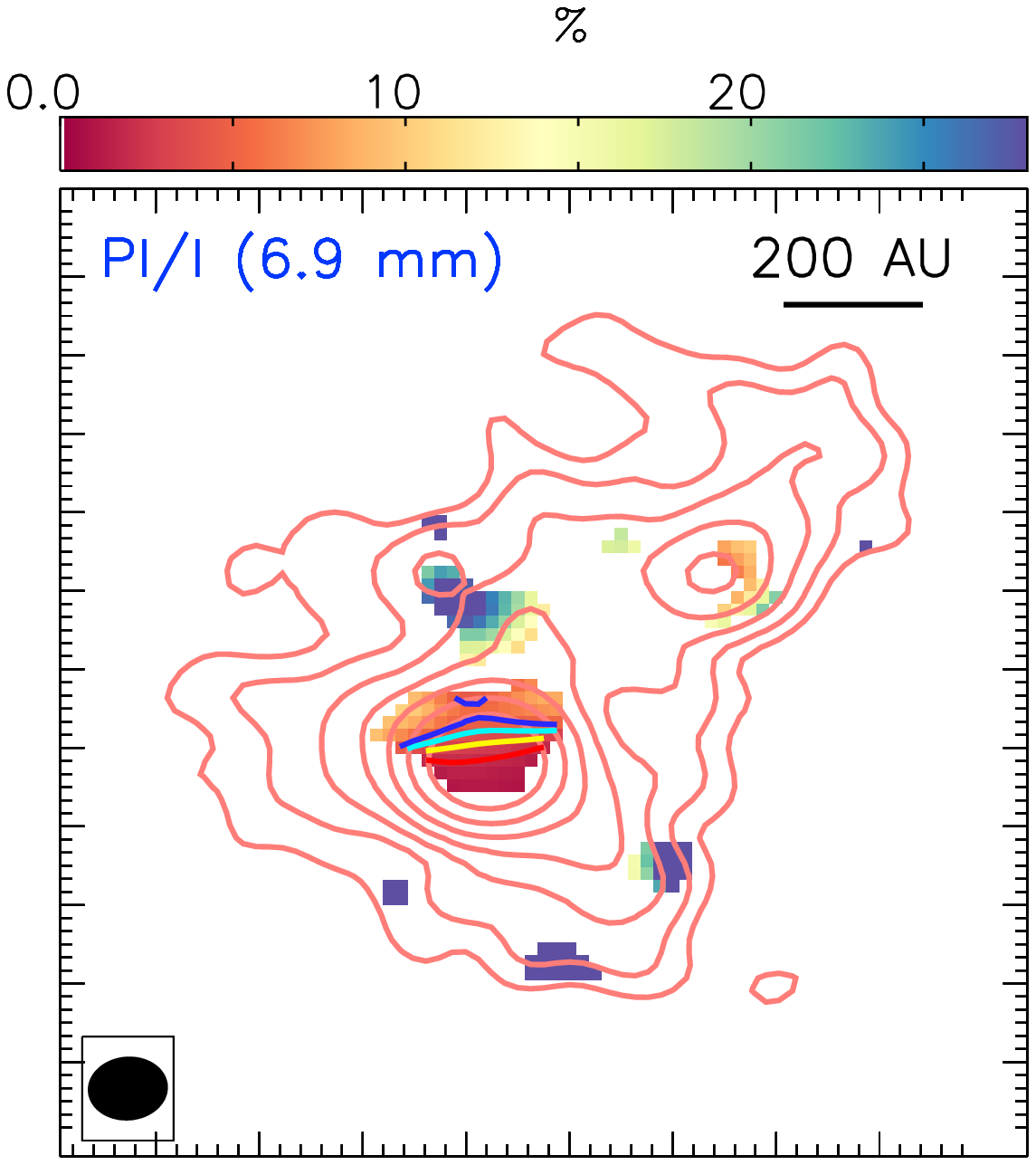} \\
\end{tabular}

\vspace{-1cm}
\hspace{-0.4cm}
\begin{tabular}{ p{4cm} p{4cm} p{4cm} p{4cm} }
\includegraphics[width=6cm]{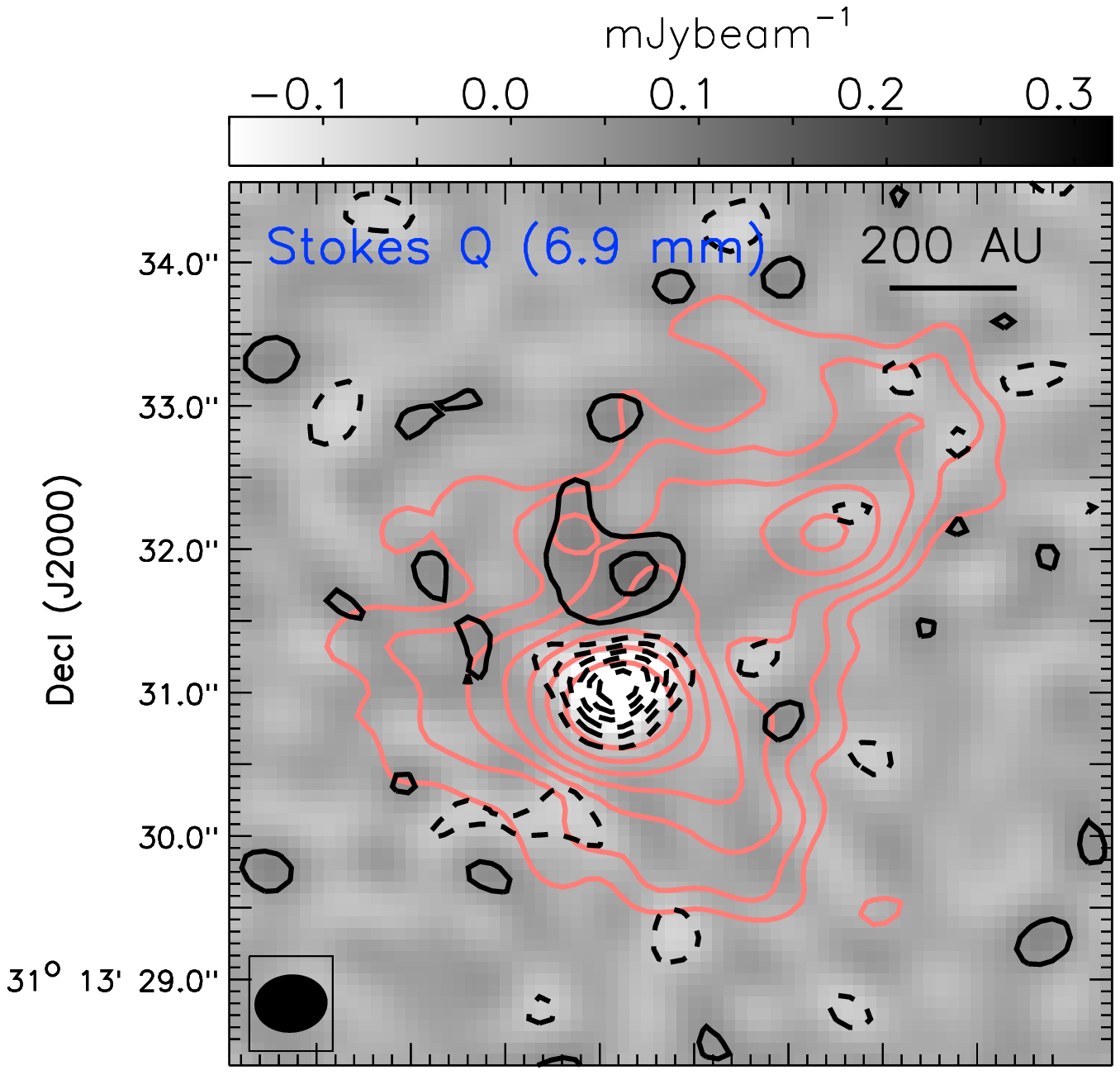} & \includegraphics[width=6cm]{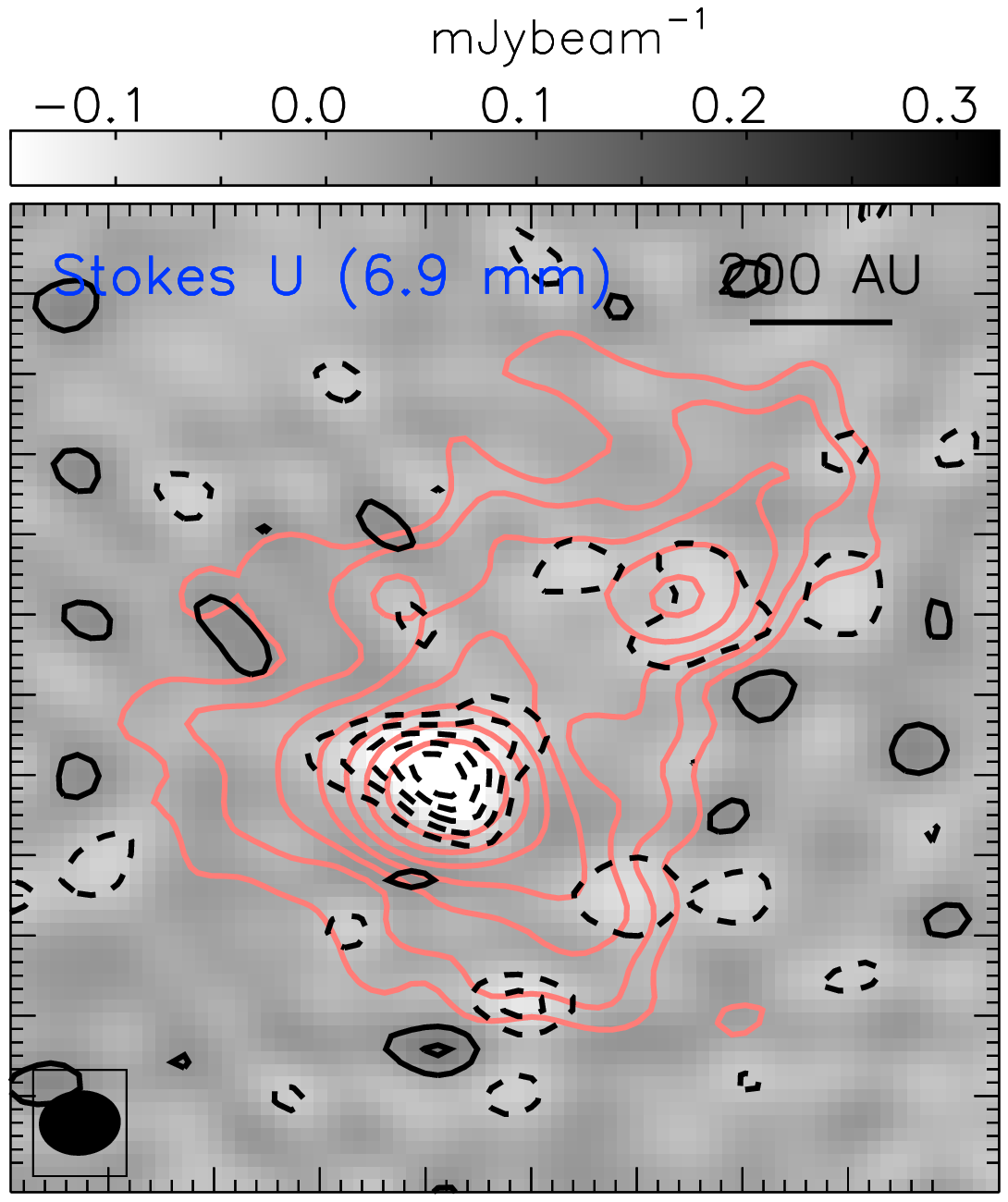} & \includegraphics[width=6cm]{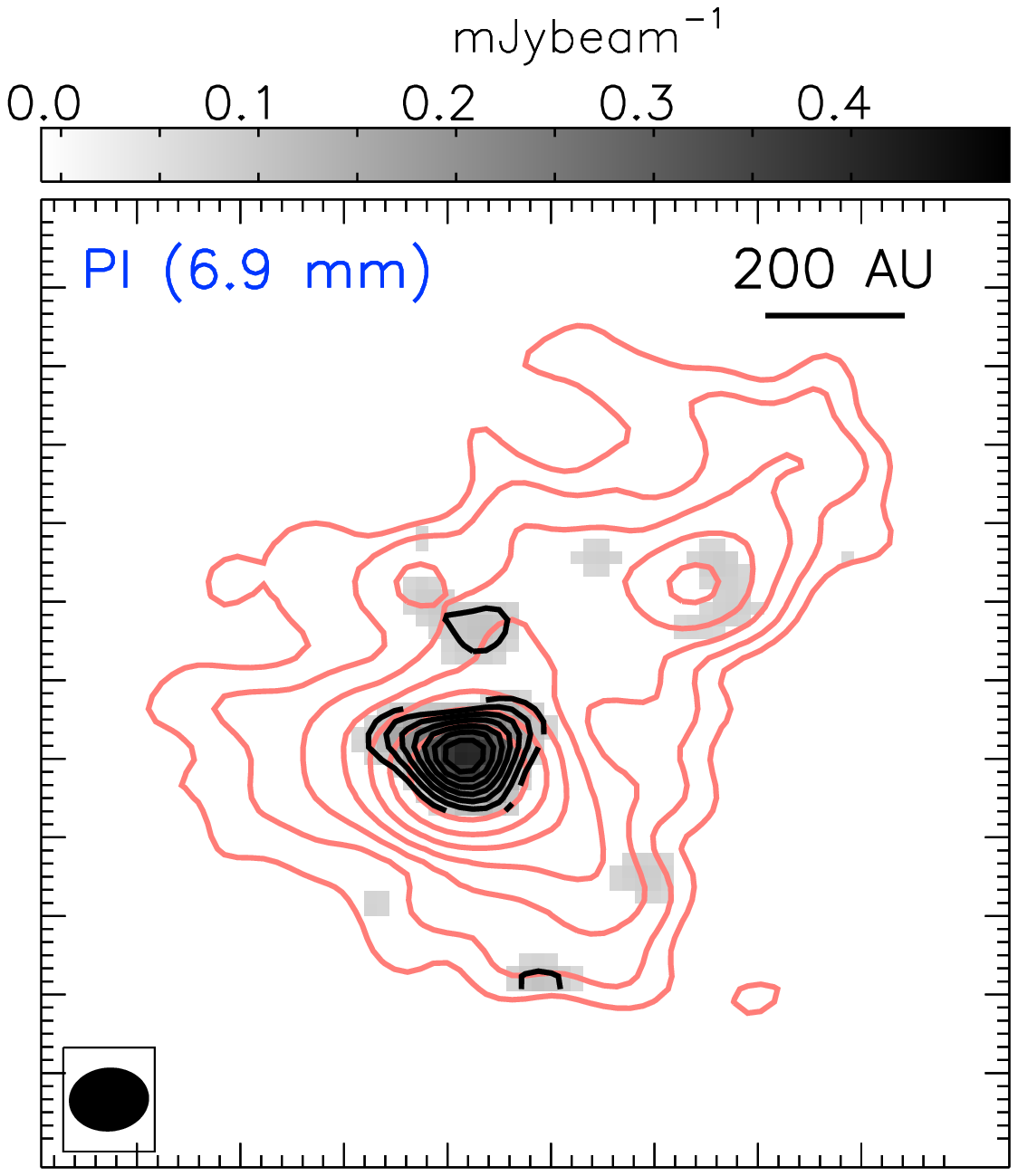}  &  \includegraphics[width=6cm]{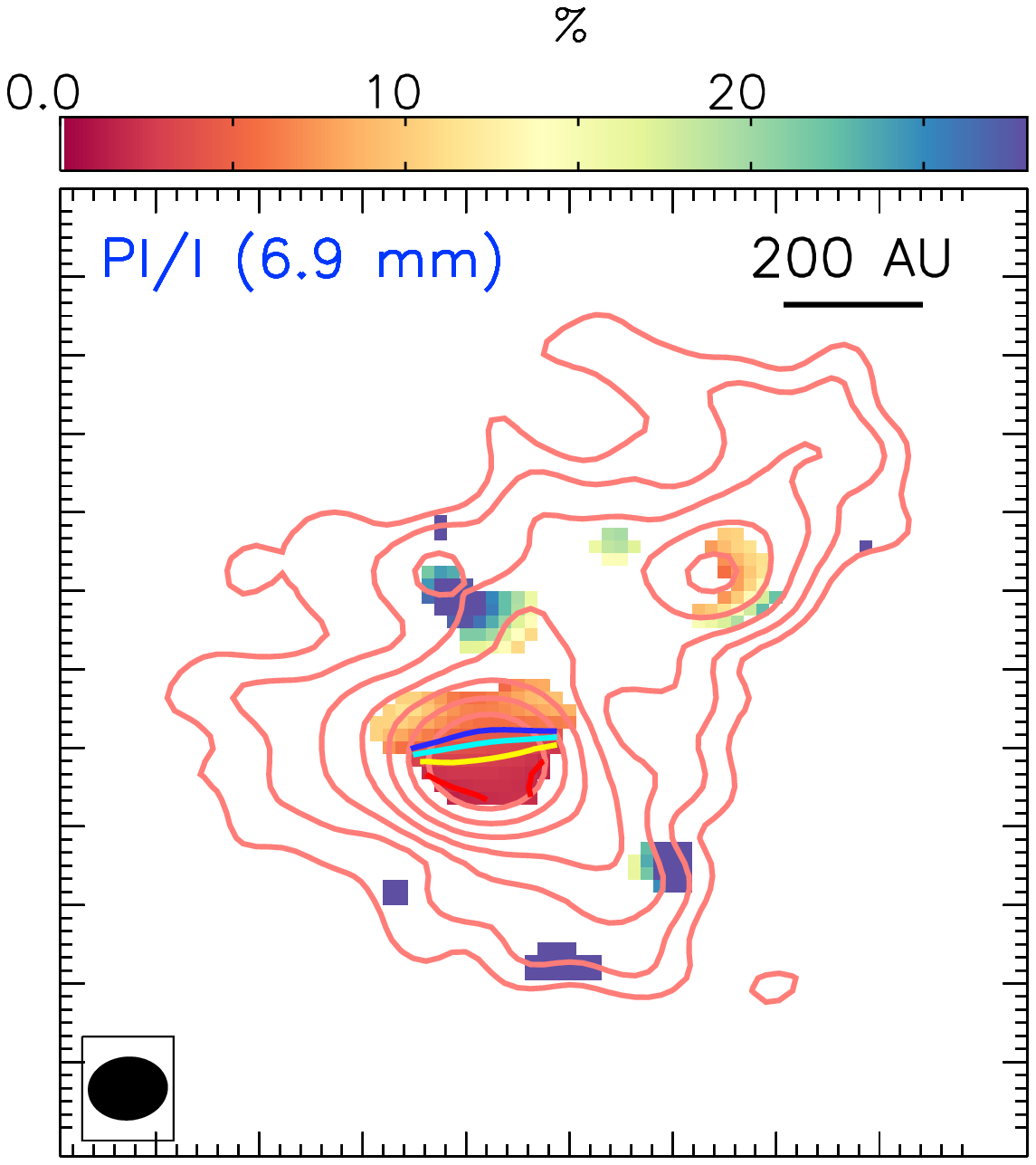} \\
\end{tabular}

\vspace{-1cm}
\hspace{-0.4cm}
\begin{tabular}{ p{4cm} p{4cm} p{4cm} p{4cm} }
\includegraphics[width=6cm]{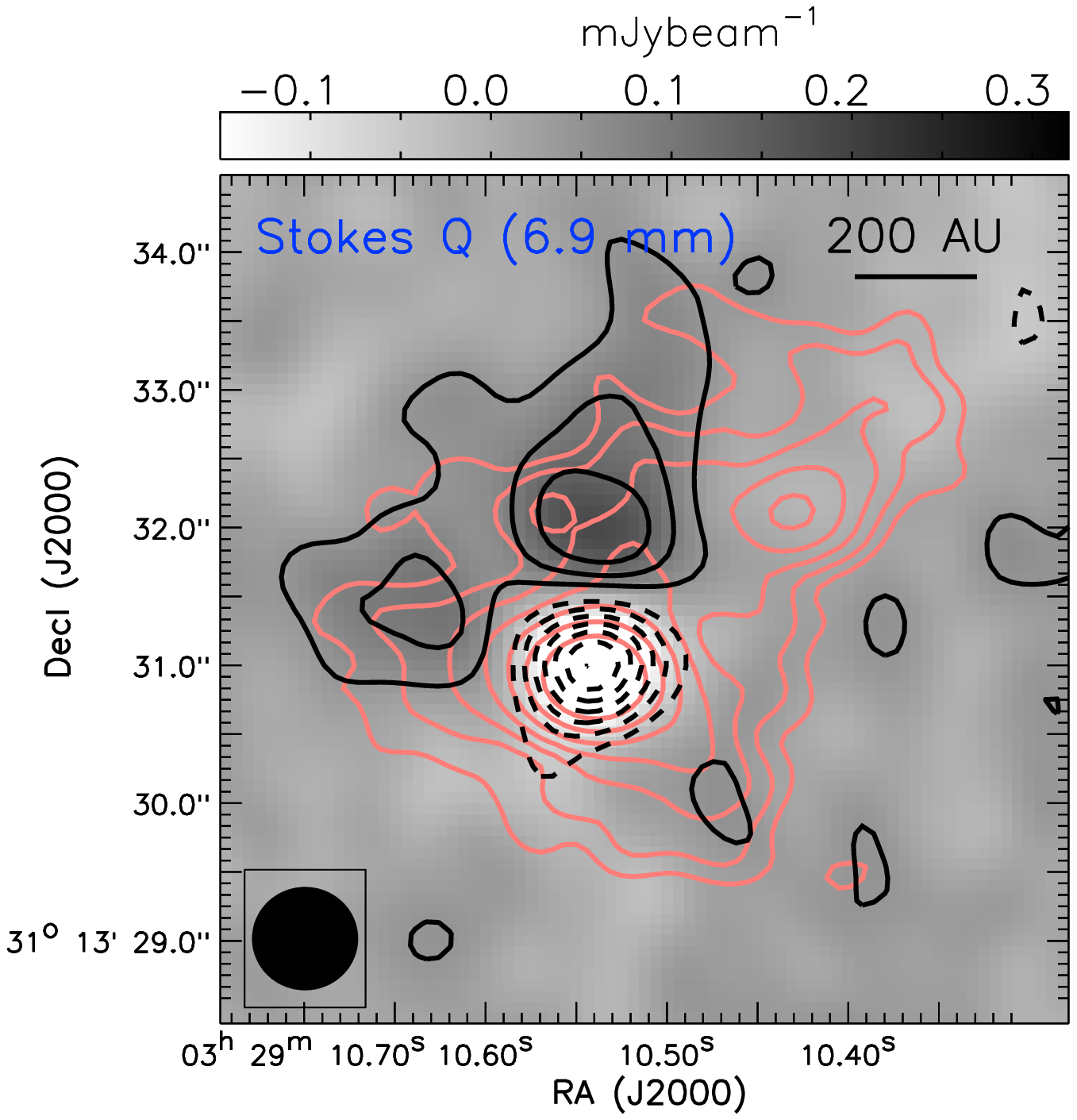} & \includegraphics[width=6cm]{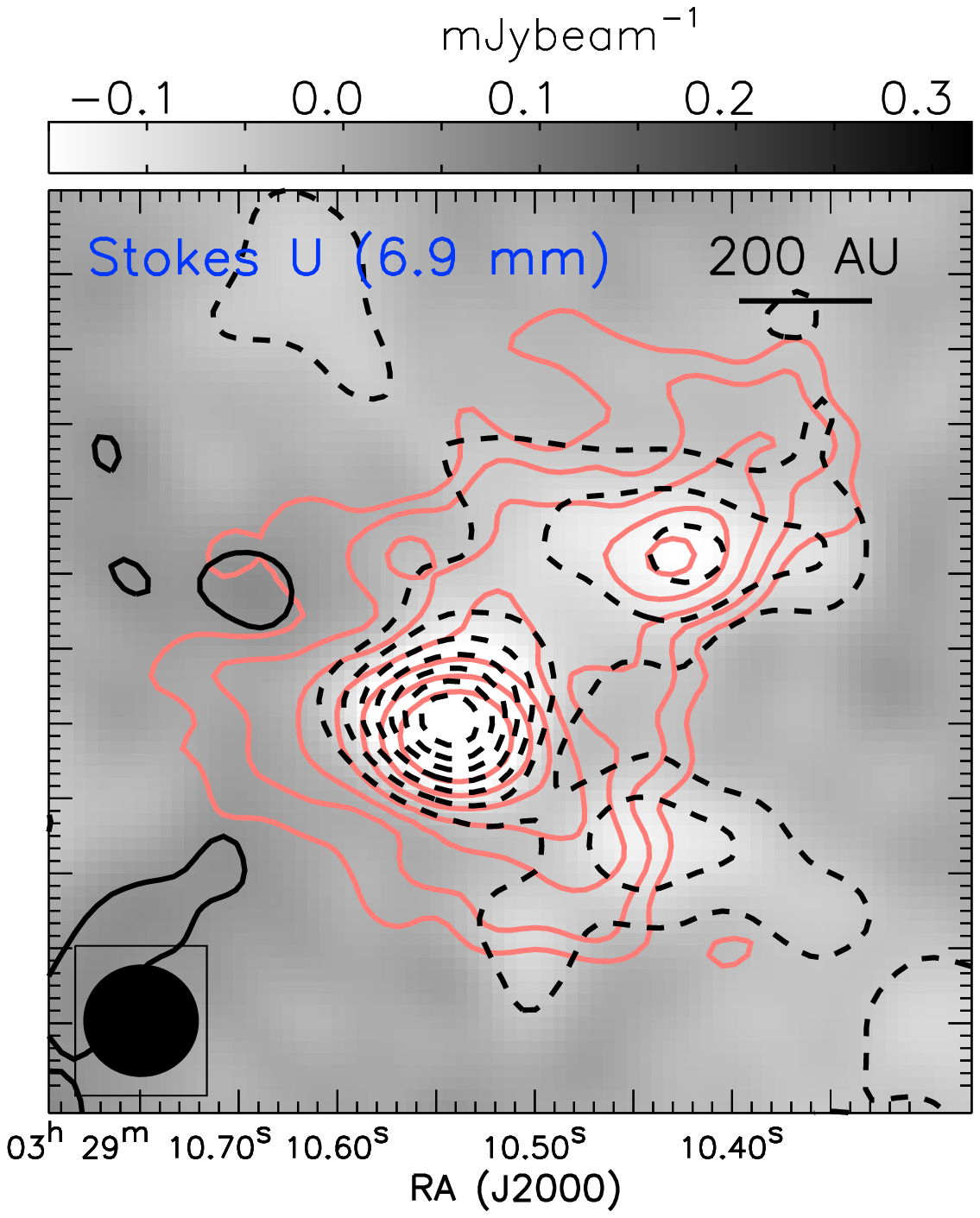} & \includegraphics[width=6cm]{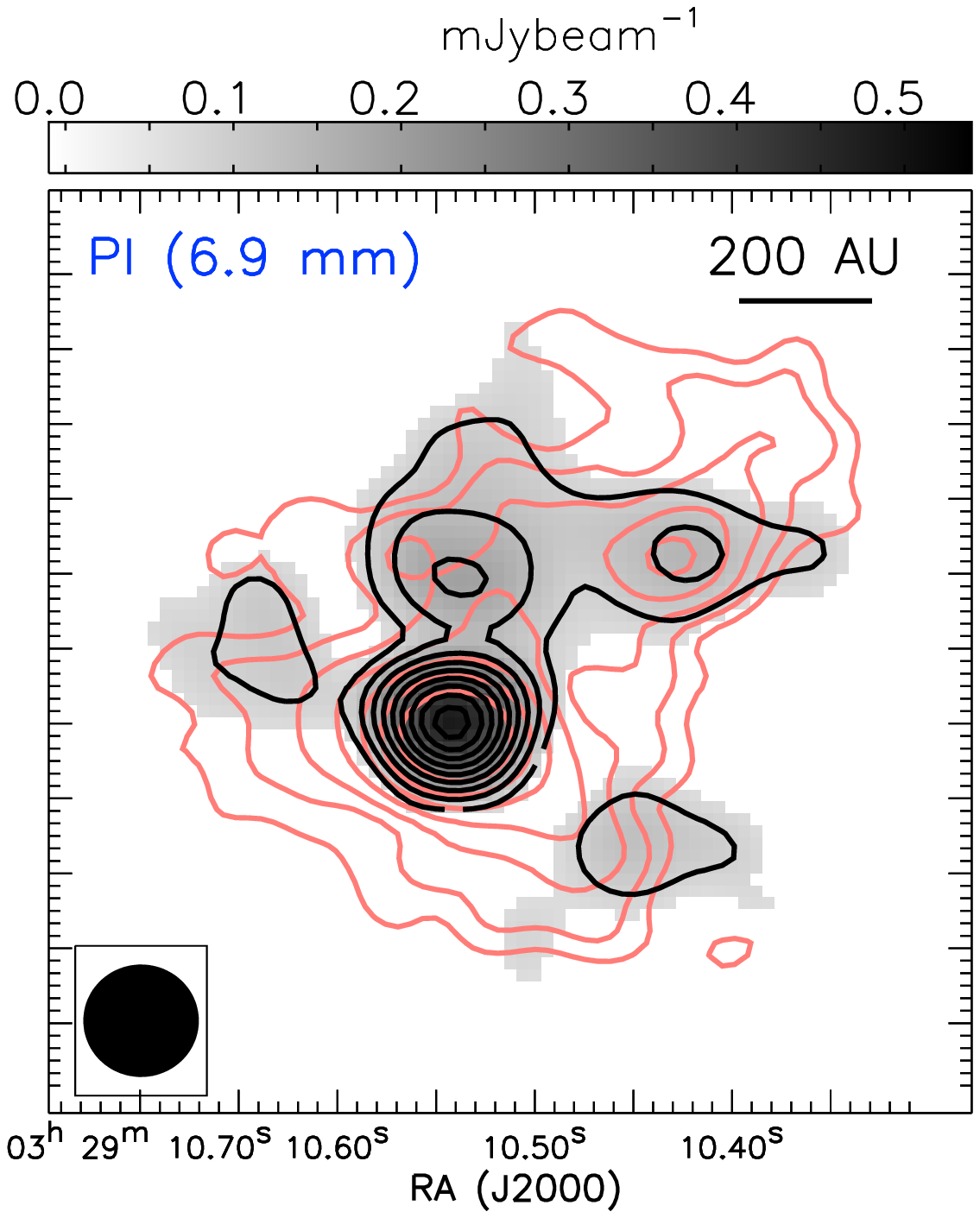}  & \includegraphics[width=6cm]{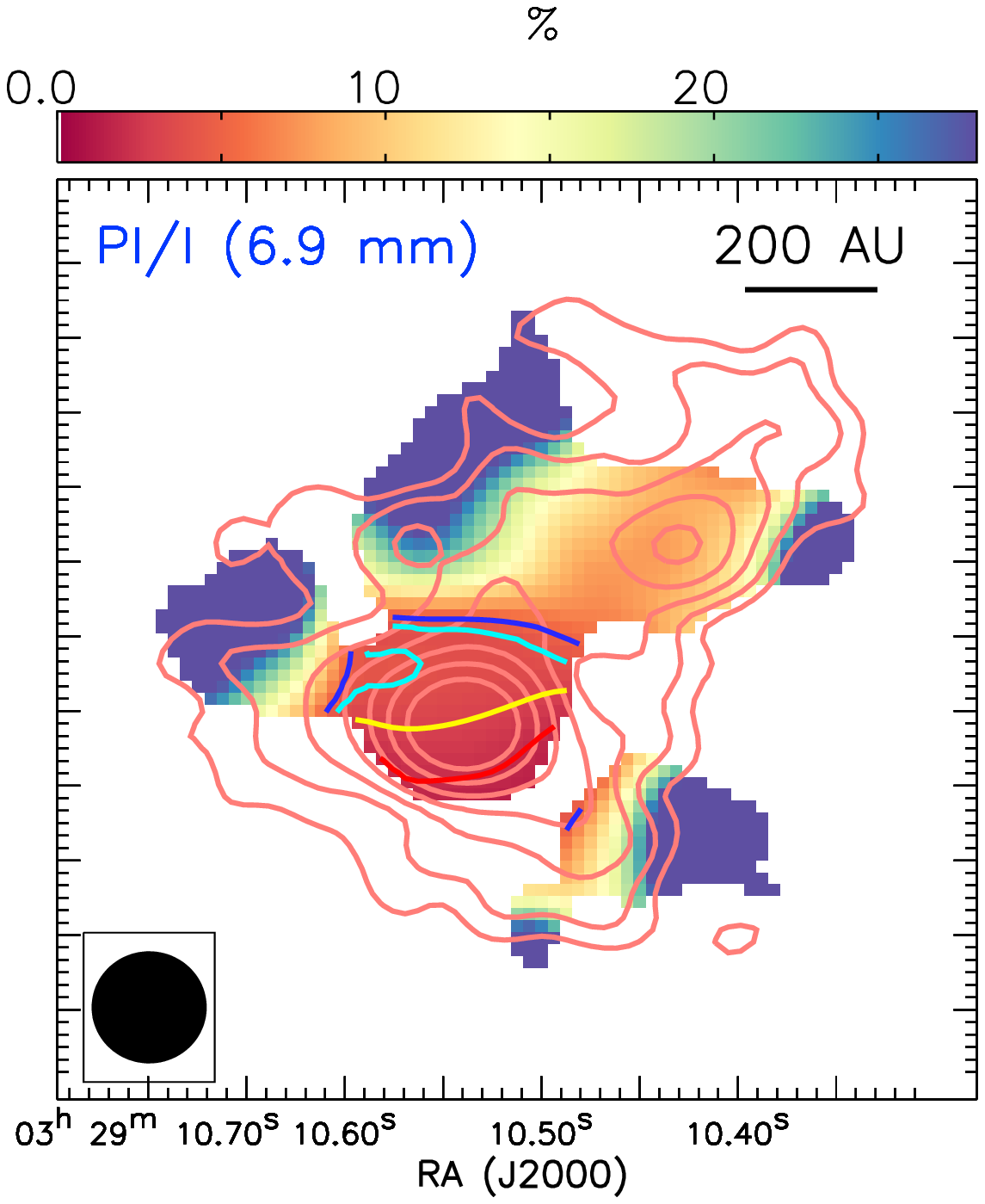}  \\
\end{tabular}
\caption{\footnotesize{
The Stokes Q (grayscale and black contours), Stokes U (grayscale and black contours), polarized intensity (PI; grayscale and black contours), and polarization percentage (PI/I; color scale and contours) images of NGC1333 IRAS4A, taken with the JVLA at 6.9 mm band.
Upper and middle rows show the JVLA images ($\theta_{\mbox{\scriptsize{maj}}}$$\times$$\theta_{\mbox{\scriptsize{min}}}$=$0\farcs49$$\times$$0\farcs40$; P.A.=-84.6$^{\circ}$) before and after removing the residual polarization leakage, respectively (for details see Section \ref{subsub:jvlapol}).
Bottom row shows the JVLA images which were generated using data in a limited $uv$ distance range of 10-455 $k\lambda$ with residual polarization leakage removed, and then were further smoothed to have an angular resolution of $\theta_{\mbox{\scriptsize{maj}}}$$\times$$\theta_{\mbox{\scriptsize{min}}}$=$0\farcs74$$\times$$0\farcs74$.
Black contours in the first three columns are 46 $\mu$Jy\,beam$^{-1}$ (2$\sigma$) $\times$[-4, -3, -2, -1, 1, 2, 3, 4, 5, 6, 7, 8, 9].
Red, yellow, light blue and dark blue contours in the last column show the 2\%, 3\%, 4\%, 5\% polarization percentage.
Orange contours in all panels show the 6.9 mm Stokes I continuum images, which have the same angular resolution and contour levels as the image presented in the left panel of Figure \ref{fig:jvlastokesI}.
The peak Stokes Q, Stokes U, and polarized intensity after removing residual polarization leakage, are -0.32 mJy\,beam$^{-1}$, -0.27 mJy\,beam$^{-1}$, and 0.40 mJy\,beam$^{-1}$, respectively in the $\theta_{\mbox{\scriptsize{maj}}}$$\times$$\theta_{\mbox{\scriptsize{min}}}$=$0\farcs49$$\times$$0\farcs40$ images.
}}
\vspace{0cm}
\label{fig:jvlaiquv}
\end{figure*}

The effects of assuming a weakly polarized source as an unpolarized calibrator in deriving the polarization leakage solutions, are well described in Sault et al. (1996).
Here we quote the immediately relevant conclusion: an error in the assumed calibrator polarization results in a spurious polarized source of the same type and percentage but of the opposite sign in the data after calibration.
In the antenna frame, these effects are identical to the full bandwidth, full Stokes images of J0713+4349, and to those of our target source NGC1333 IRAS4A.
Therefore, to zeroth order, we are able to remove the residual polarization leakage from the full Stokes images of our target source in the antenna frame, by
\begin{equation}
Q_{target} = Q'_{target} - I'_{target}\times\frac{Q'_{J0713+4349}}{I'_{J0713+4349}},
\label{eq:1}
\end{equation}
and 
\begin{equation}
U_{target} = U'_{target} - I'_{target}\times\frac{U'_{J0713+4349}}{I'_{J0713+4349}},
\label{eq:2}
\end{equation}
where $I_{i}$, $Q_{i}$, $U_{i}$ are the Stokes I, Q, and U fluxes of the source $i$ after the residual polarization leakage removal, and the primed ones are the fluxes before the residual polarization leakage removal.
During our observations, our target source NGC1333 IRAS4A nearly did not rotate with respect to the receivers. 
The parallactic angle range of NGC1333 IRAS4A is 74.5$^{\circ}$$\pm$1.5$^{\circ}$.
Therefore, we neglected the effect of field rotation during the target source scans.
We note that assuming the measured $Q'_{J0713+4349}/I'_{J0713+4349}$ and $U'_{J0713+4349}/I'_{J0713+4349}$ are small, it is possible to analytically derive the correction terms for the polarization leakage solutions.
Corrections were made in the image domain.
In the case that the correction terms are small ($\sim$1\%), the subtractions in Equations \ref{eq:1} and \ref{eq:2} will not impact the thermal noise level.

J0713+4349 shows 1.2\% polarization percentage with a $\sim$90$^{\circ}$ polarization position angle  after the calibration steps introduced in Section \ref{subsub:jvlacal}.
Strictly speaking, these values are the combined effects of the weak polarization of 3C84 when deriving the polarization leakage solution, and the intrinsic polarization of J0713+4349.
The J0713+4349 polarization is usually less than 0.1\% according to the Guide to Observing with the VLA, despite some peaks of polarization percentage of up to 2.69\%  (Table \ref{tab-polcal}). 
We do not know the exact polarization percentage of J0713+4349 at the time when we observed it.
We quote the latest polarization percentage peak of $\sim$0.3\% as an upper limit of our final polarization calibration error after the removal of residual polarization leakage.
Considering the observed peak intensity of Stokes I continuum emission of 12.7 mJy\,beam$^{-1}$ (Section \ref{subsub:jvlacal}), the maximum of the spuriously polarized intensity due to our final polarization calibration error is 38.1 $\mu$Jy\,beam$^{-1}$ (1.7$\sigma$).
Subject to this effect, the observed polarization position angle at the location of Stokes I continuum peak may be rotated, and the polarization percentage may be biased.
Outside of the one synthesized beam area surrounding the peak of Stokes I continuum emission, the spuriously polarized signal is unlikely to be detected given our present sensitivity.
We caution that our observational setup does not allow us to characterize the off-center polarization leakage.
Nevertheless, given that the region of our interest is merely in the central $<$4$''$ area, which is $\sim$10 times smaller than the full width at half maximum of the antenna primary beam, we expect the polarization leakage to be approximately identical across this region.
The peak Stokes Q, Stokes U, and polarized intensity after removing residual polarization leakage, are -0.32 mJy\,beam$^{-1}$, -0.27 mJy\,beam$^{-1}$, and 0.40 mJy\,beam$^{-1}$, respectively.

% list our assumptions: we assume the values of Q/I and U/I of 3C84 have weak dependence on frequency -> check it by looking at the polarization property of J0713+4349.

% Check the J0713+4349 results by imaging the first 32 and last 32 spws, and see if we get consistent Stokes Q and U

\begin{figure}
\hspace{-1cm}
\begin{tabular}{p{4cm} p{4cm}}
\includegraphics[width=6cm]{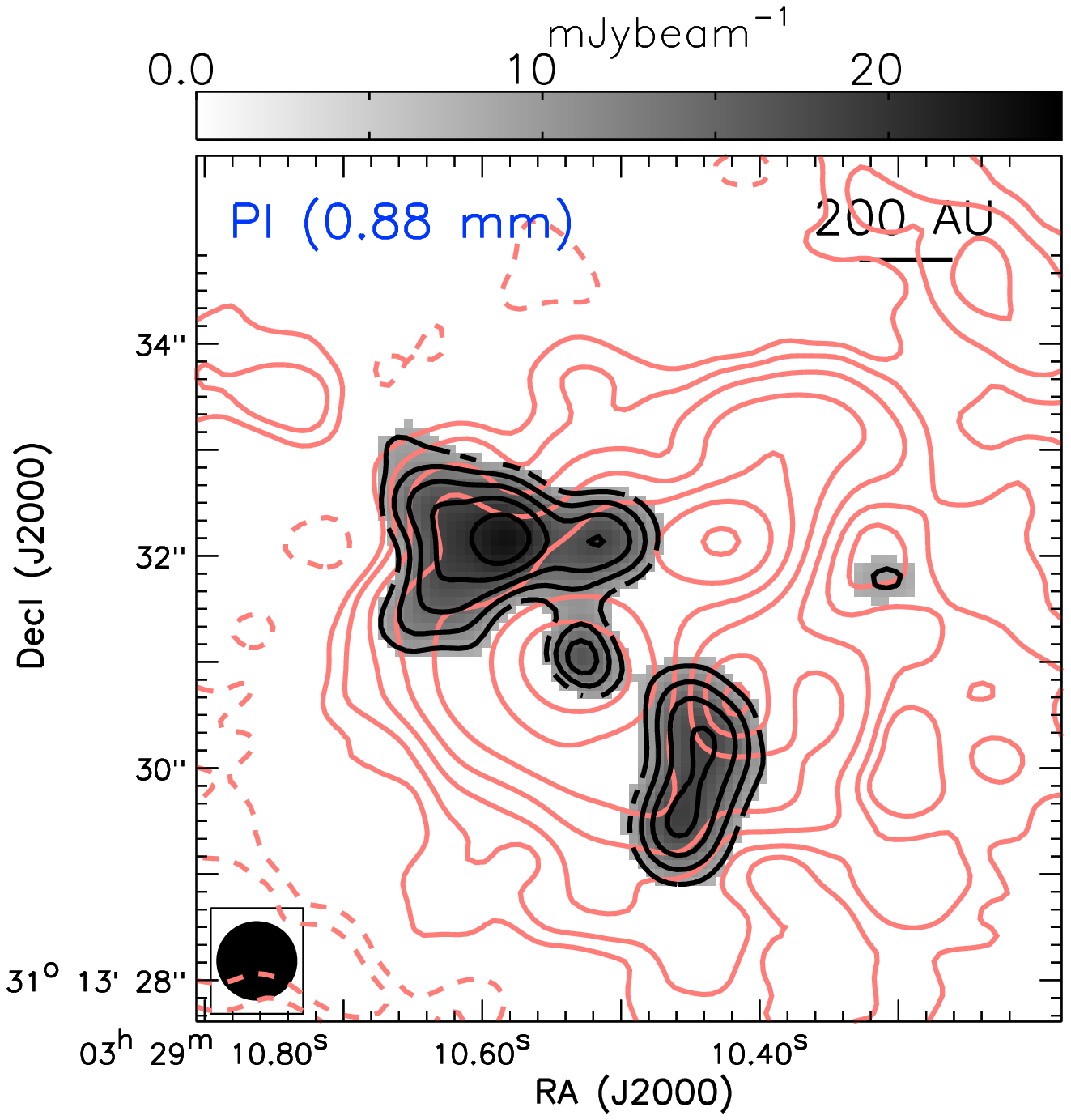} & \includegraphics[width=6cm]{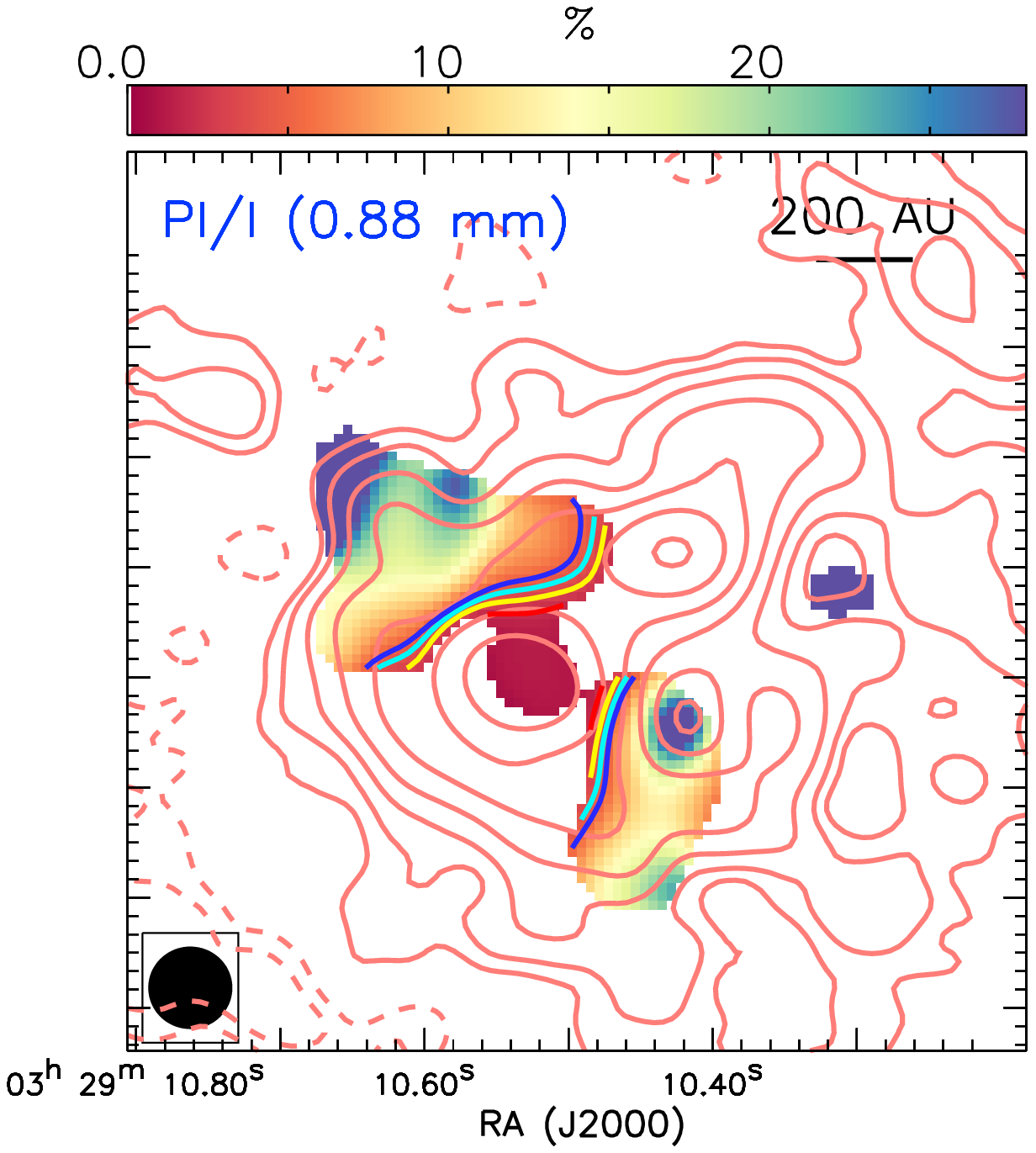}\\
\end{tabular}
\vspace{0cm}
\caption{\footnotesize{
Results of the SMA 0.88 mm polarization observations ($\theta_{\mbox{\scriptsize{maj}}}$$\times$$\theta_{\mbox{\scriptsize{min}}}$=$0\farcs74$$\times$$0\farcs74$). Gray and color scales show the polarization intensity and the polarization percentage images, respectively. Orange contours show the 0.88 mm Stokes I  continuum emission. Black contours show the polarized intensity at 0.88 mm. Orange contours are 9 mJy\,beam$^{-1}$ (6$\sigma$) $\times$[-2, -1, 1, 2, 4, 8, 16, 32, 64]. Black contours are 3 mJy\,beam$^{-1}$ (2$\sigma$) $\times$[1, 2, 3, 4, 5]. Red, yellow, light blue and dark blue contours in the right panel show the 2\%, 3\%, 4\%, 5\% polarization percentage.
}}
\vspace{0.5cm}
\label{fig:sma}
\end{figure}

\begin{figure}
\hspace{-0.6cm}
\begin{tabular}{p{8.8cm} }
\includegraphics[width=9.35cm]{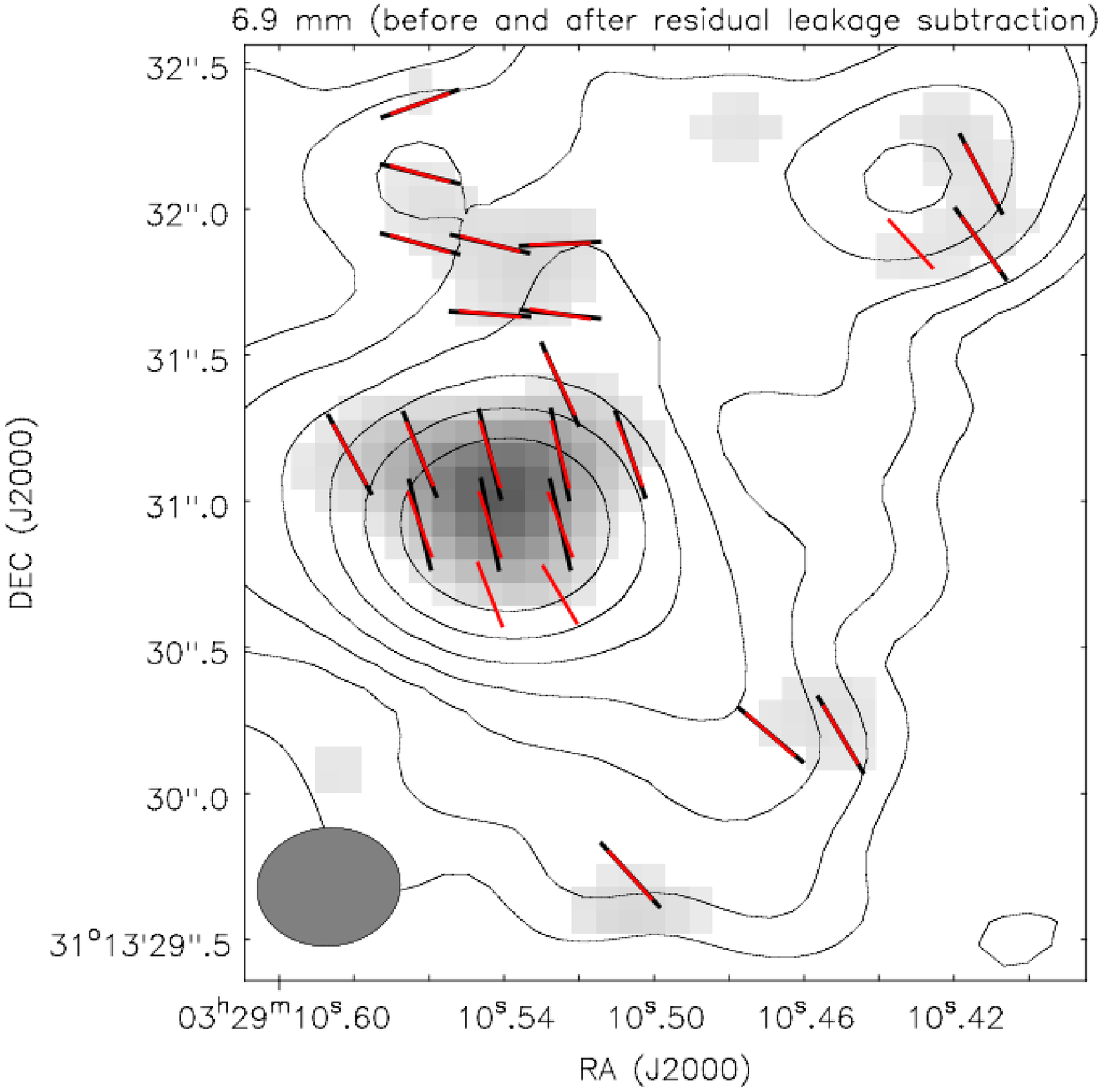} \\
\vspace{-3.5cm}
\includegraphics[width=9.35cm]{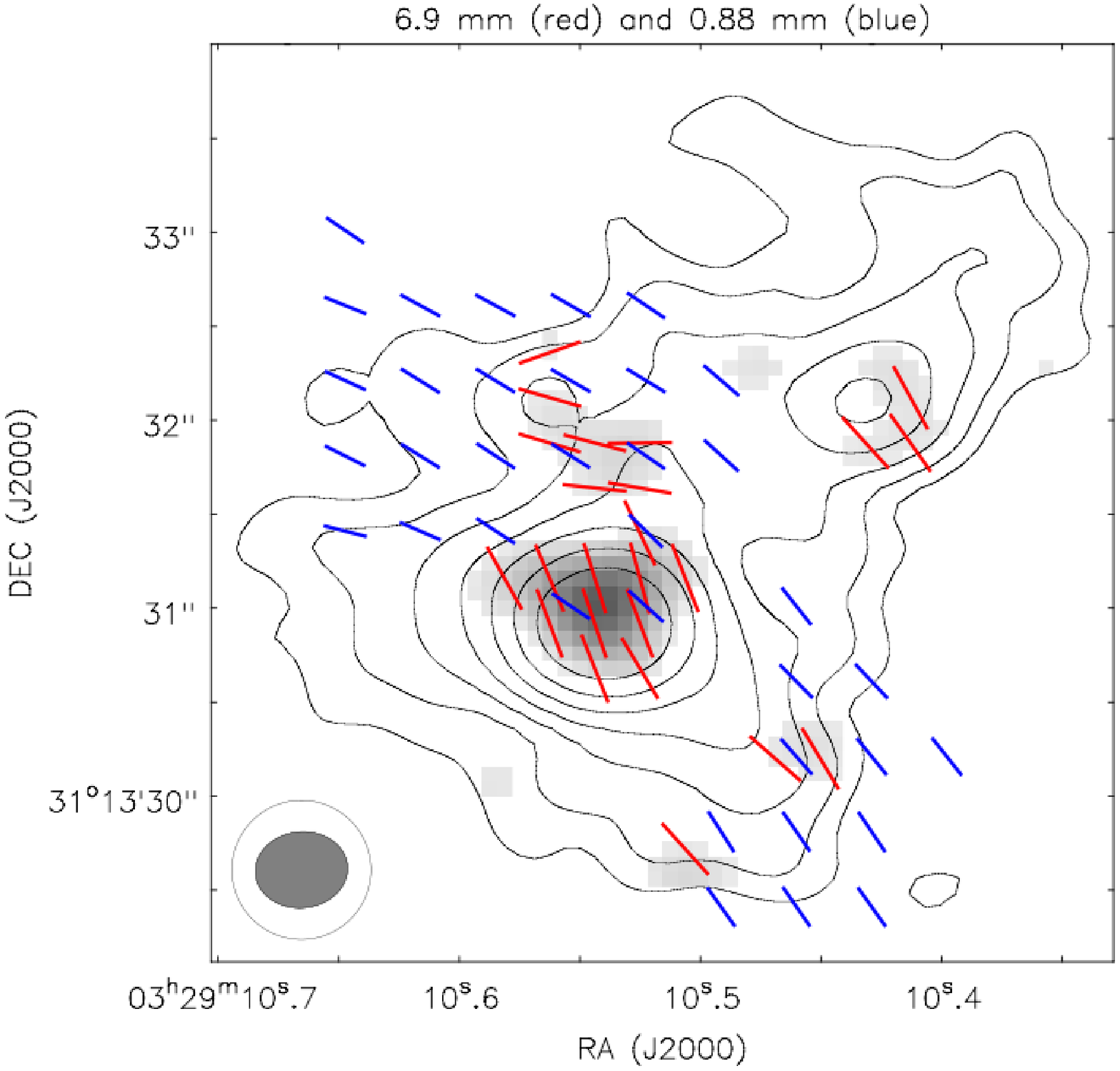}\\
\end{tabular}
\vspace{-4.5cm}
\caption{\footnotesize{
The observed B-segment position angle (i.e. 90 degree rotated from the E-field orientation) at 7mm by JVLA, and at 0.88 mm by SMA.
Top panel shows the comparison of the JVLA observations, before residual leakage removal (black), and red after residual leakage removal (red).
Bottom panel shows the comparison of the JVLA observation (after leakage removal; red) with the SMA observations (blue).
The SMA data presented in this panel were generated using data in the $uv$ distance range of 10-455 $k\lambda$.
Contours present the 6.9 mm Stokes I continuum emission.
The polarized intensity at 6.9 mm is presented in grayscale (see Figure \ref{fig:jvlaiquv}).
Contours are 69 $\mu$Jy\,beam$^{-1}$ (3$\sigma$) $\times$ [-2, -1, 1, 2, 4, 8, 16, 32, 64].
The filled and open ellipses in the bottom left show the synthesized beams of the JVLA and the SMA image, respectively.
The observed polarization position angles before and after residual leakage removal show up to $\sim$10$^{\circ}$ differences at the center of IRAS4A1. The differences become negligibly small in the outer region.
}}
\vspace{0.25cm}
\label{fig:angle}
\end{figure}

\subsection{The SMA Data}
\label{sub_smadata}
% com  : 2004 Dec 05, 06 (lsb: 335.55618 GHz; 10~165 kl)
% sub   : 2009 Jan 27 (lsb: 336.19160 GHz; 10~80 kl)
% ext   : 2009 Feb 24 (lsb: 336.18977 GHz; 30~260 kl)
% vex: 2011 Jan 24 (lsb: 333.5-337.4; usb: 345.5-349.4), 2011 Jan 25 (lsb: 333.5-337.4; usb: 345.5-349.4; 70-520 kl,)
The SMA compact array data were taken on 2004 December 05 and 06, covering the $uv$ distance range of 10-165 $k\lambda$.
The subcompact array data were taken on 2009 January 27, which covered the $uv$ distance range of 10-80 $k\lambda$.
The extended array data were taken on 2009 February 24, covering the $uv$ distance range of 30-260 $k\lambda$.
The very extended array data were taken on 2011 January 24 and 25, which covered the $uv$ distance range of 70-520 $k\lambda$.
All these SMA observations were centered at $\sim$340 GHz (0.88 $\mu$m), and included the full RR, RL, LR, and LL correlator products. 
Bright quasars were tracked with large parallactic angle coverages for deriving the polarization leakage solutions. 

Data calibrations and imaging were carried out using the Miriad software package (Sault et al. 1995). 
We refer to Girart et al.(2006) for results of the compact array observations.
We refer for more details of these SMA observational results to Frau et al. (2011), and Ching et al. (2016).
For the present paper, the SMA data within the $uv$ distance range of 10-455 $k\lambda$ were jointly imaged, to ensure the more fair comparison between the JVLA observations (Section \ref{sub_jvladata}) and the SMA data in the same $uv$ distance range.
We note that selecting the same $uv$ distance range does not guarantee the same brightness sensitivities on all sampled angular scales, which are related to the detailed distributions of the $uv$ sampling points of the JVLA and the SMA data.
The achieved RMS noise level in the SMA Stokes I, Q, and U images are 4.37, 1.5, and 1.5 mJy\,beam$^{-1}$, respectively.
The Stokes I image is affected by limited dynamical range and by the missing flux at the short visibility spacings.
As a consequence, the Stokes I RMS noise level is 3 times higher than that of the Stokes Q and U images.
The synthesized beam of the restored images is $\theta_{\mbox{\scriptsize{maj}}}$$\times$$\theta_{\mbox{\scriptsize{min}}}$=$0\farcs74$$\times$$0\farcs61$ (P.A.=78$^{\circ}$).
We further smoothed the SMA images to have circular synthesized beam of $\theta_{\mbox{\scriptsize{maj}}}$$\times$$\theta_{\mbox{\scriptsize{min}}}$=$0\farcs74$$\times$$0\farcs74$.
% (Some information about uv-distance range, polarization leakage, and RMS sensitivity here. Details refer to Shih-Ping paper)

\begin{figure}
\hspace{-0.6cm}
\begin{tabular}{p{8.8cm}}
\includegraphics[width=9.35cm]{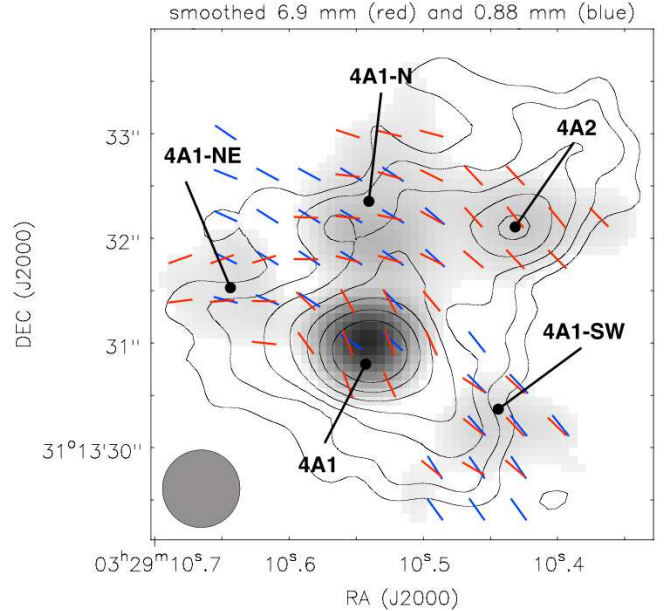}\\
\end{tabular}
\vspace{-1.1cm}
\caption{\footnotesize{
The comparison of observed B-segment position angle (i.e. 90 degree rotated from the E-field orientation) at 7mm by JVLA (red), and at 0.88 mm by SMA (blue).
The JVLA and SMA images presented in this panel were generated using the data in $uv$ distance range of 10-455 $k\lambda$.
Both JVLA and SMA images were smoothed to have circular beams of $\theta_{\mbox{\scriptsize{maj}}}$$\times$$\theta_{\mbox{\scriptsize{min}}}$=$0\farcs74$$\times$$0\farcs74$.
Contours present the 6.9 mm Stokes I continuum emission.
The polarized intensity at 6.9 mm is presented in grayscale (see Figure \ref{fig:jvlaiquv}).
Contours are 69 $\mu$Jy\,beam$^{-1}$ (3$\sigma$) $\times$ [-2, -1, 1, 2, 4, 8, 16, 32, 64].
}}
\vspace{0.3cm}
\label{fig:angle_smooth}
\end{figure}

\subsection{Polarization Images}
\label{sub_polimage}
After generating the Stokes I, Q and U images from the JVLA and the SMA data, we produced the polarization intensity images, the polarization percentage images, and the polarization position angle images using the Miriad task {\tt impol}.
The derived polarization intensity images are de-biased according to $P$=$\sqrt{Q^2 + U^2 - \sigma_P^2}$, where P is the expected actual polarized intensity, and $\sigma_P$ is the RMS noise of $\sqrt{Q^2+U^2}$ (for more discussion see Vaillancourt 2006, Hull et al. 2014).
We implemented a 3$\sigma$ cutoff for the Stokes I, Q and U images to suppress the positive bias in the polarization intensity images, and to avoid the spuriously polarized intensity due to the polarization calibration error (see Section \ref{subsub:jvlapol}).
However, the derived polarization percentages might still be biased to higher values in extended regions, because the Stokes I images are more subjected to missing short spacing data.
The polarization percentages at 6.9 mm and 0.88 mm can be better constrained in the future, after incorporating the James Clerk Maxwell Telescope (JCMT) or Atacama Pathfinder EXperiment (APEX) 0.88 polarization observations, and the JVLA 6.9 mm observations in the more compact (D) array configuration (see Girart et al. 2006 for discussion).

%============================================================
% uv-distance range for comparing SMA and JVLA results: 10-455 klambda
\section{Results}
\label{chap_result}

\begin{figure}
\vspace{-0.4cm}
\begin{tabular}{p{5cm} }
\includegraphics[width=8.5cm]{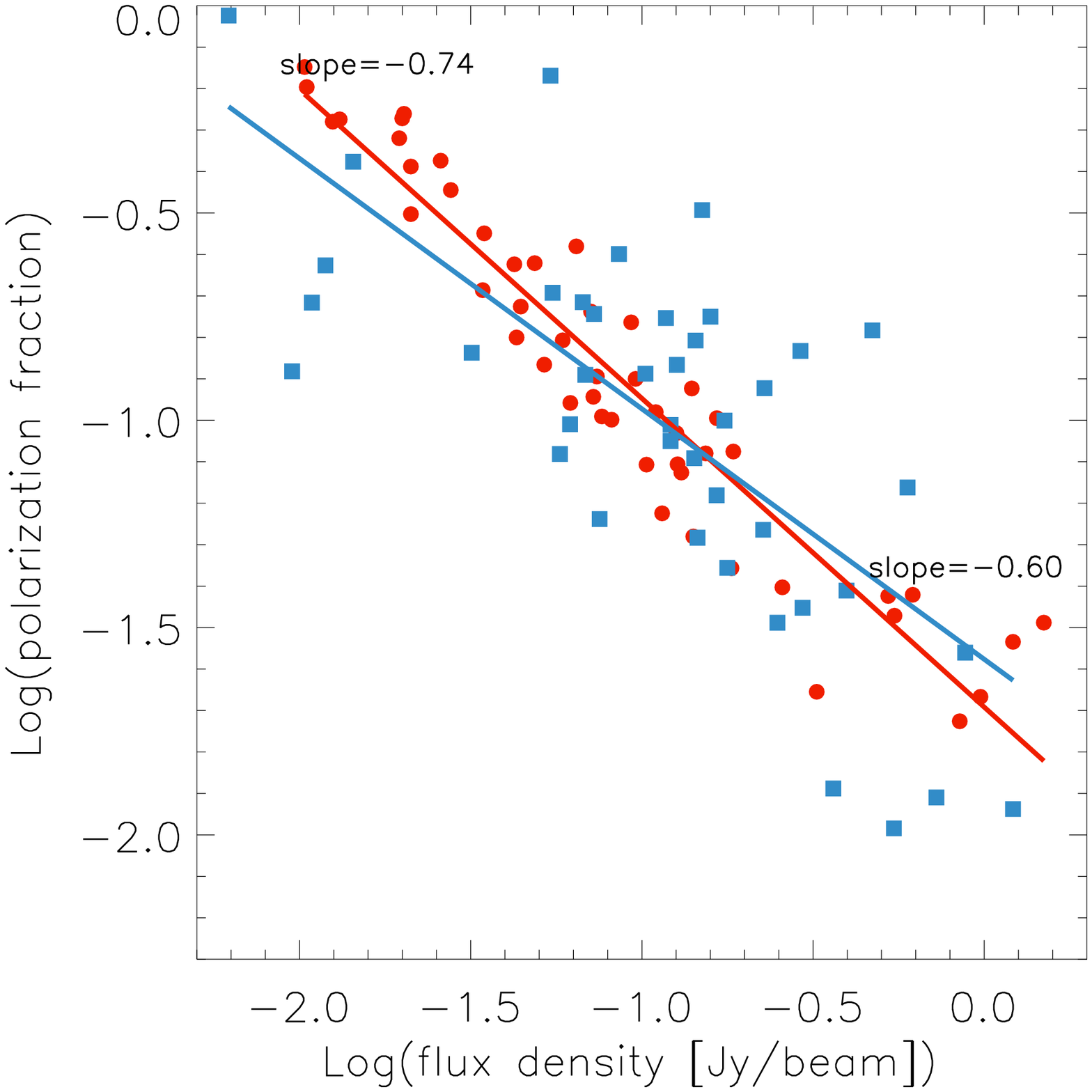} \\
\includegraphics[width=8.5cm]{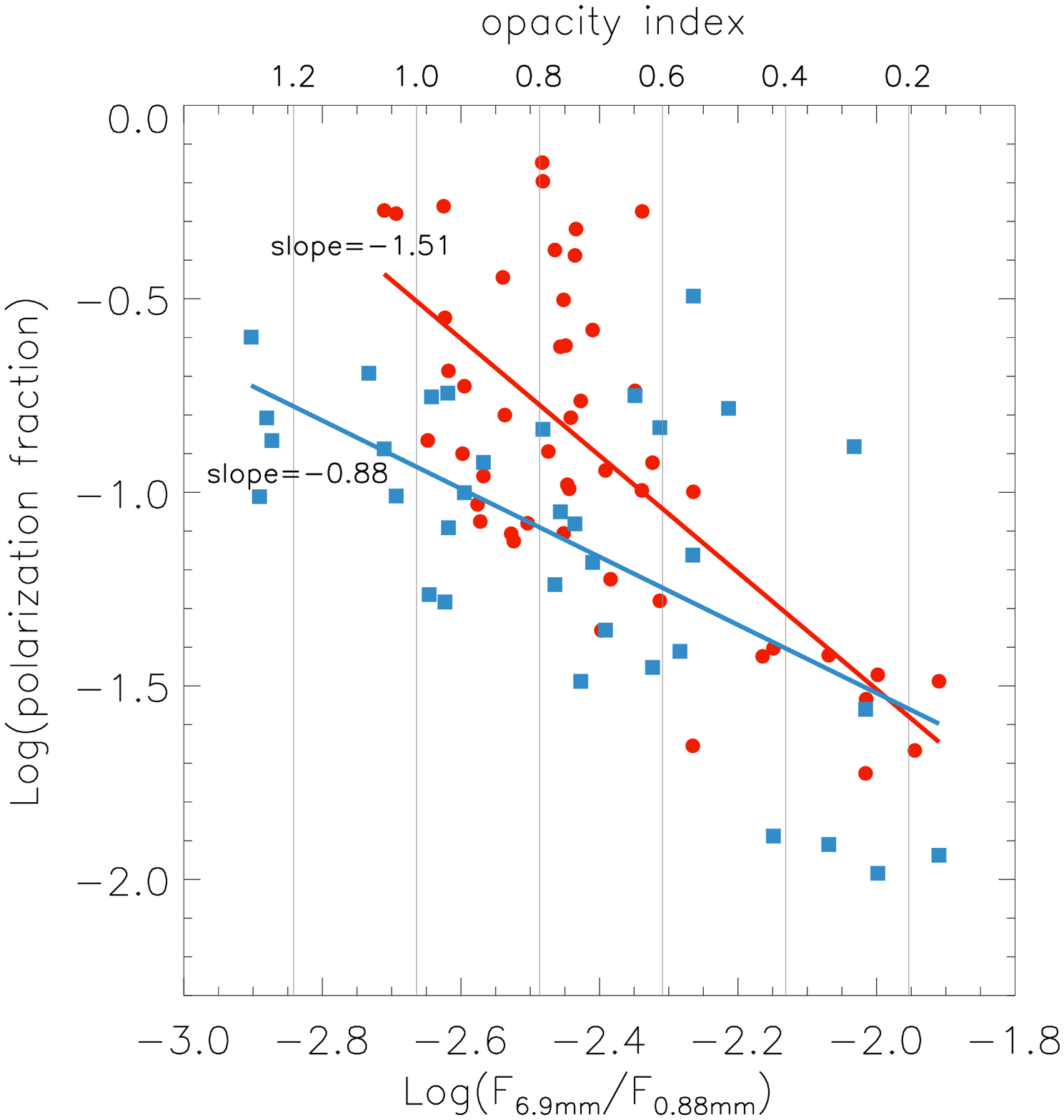} \\
\end{tabular}
\caption{\footnotesize{
Comparison of the polarization percentage observed by JVLA at 6.9 mm (red), and the polarization percentage observed by SMA at 0.88 mm (blue). 
Top panel plots polarization percentages as a function of Stokes I continuum flux densities at 6.9 mm and 0.88 mm bands; bottom panel plots polarization percentages as a function of the flux ratio of Stokes I continuum emission at 6.9 mm and 0.88 mm.
Sampling is made approximately twice per beam size in each dimension.
The JVLA and SMA images presented in these panels were generated using data in the $uv$ distance range of 10-455 $k\lambda$.
Both JVLA and SMA images were smoothed to have circular beams of $\theta_{\mbox{\scriptsize{maj}}}$$\times$$\theta_{\mbox{\scriptsize{min}}}$=$0\farcs74$$\times$$0\farcs74$.
The 6.9 mm flux densities in the bottom panel are multiplied by a factor of 100.
Red and blue lines show linear regressions for the plotted SMA and JVLA data points.  
Due to missing the extended fluxes, the obtained slopes from linear regressions are lower limits. 
Gray lines in the lower panel indicate the corresponding dust opacity spectral index ($\beta$) of the detected flux ratios, which were estimated based on the Rayleigh-Jeans assumption (i.e. $F_{\nu}$$\propto$$\nu^{2+\beta}$).
Overall this target source shows the higher 6.9 mm to 0.88 mm flux ratios towards the brighter (sub)millimeter emission regions.
The observed polarization percentages are steeply decreasing with increasing 6.9 mm to 0.88 mm flux ratios as well as with increasing (sub)millimeter intensities.
}}
\label{fig:diagrams}
\end{figure}

All images presented in this paper were corrected for the antenna primary beam responses, if not specifically mentioned. 
Figure \ref{fig:jvlastokesI} shows Stokes I images of all the sources detected at 6.9 cm in our JVLA observations.
NGC1333 IRAS4A presents an elongated, 900 AU scale structure, which is harboring the previously known Class 0 YSOs, IRAS4A1 and IRAS4A2 (e.g. Girart et al. 2006; Choi et al. 2011).
The protobinary IRAS4B and IRAS4B2 are also detected $\sim$30$''$ away from the phase referencing center, which is close to the edge of the primary beam.
IRAS4B is partially resolved along the southeast-northwest direction.
% We note that on large scales, the direction of B-field is approximately perpendicular to the elongated structure in IRAS4A (Girart et al. 2006; Matthews et al. 2009; Hull et al. 2014).

The linearly polarized continuum emission is only detected around the IRAS4A region in both JVLA and SMA images.
The 6.9 mm Stokes Q, U, polarization intensity, and polarization percentage images, are shown in Figure \ref{fig:jvlaiquv}.
Figure \ref{fig:sma} shows the 0.88 mm Stokes I, polarization intensity, and polarization percentage images taken with the SMA.
The inferred B-field orientations (referred to as B-segments hereafter) are plotted in Figure \ref{fig:angle} and \ref{fig:angle_smooth}, and have been rotated by 90 degrees from the measured E-field orientations.
% We refer to the 90$^{\circ}$ rotated E field polarization orientations {\it as B-segments}, hereafter.
% From Figure \ref{fig:angle}, we see that after removing the residual polarization leakage (see Section \ref{subsub:jvlapol}), the observed polarization position angles by JVLA at the peak of 6.9 mm polarization intensity, agrees with those observed by SMA.
At the location where the polarization percentage before removing residual polarization leakage is comparable to 3\% or smaller (c.f. Figure \ref{fig:jvlaiquv}), our strategy of removing residual polarization leakage (see Section \ref{subsub:jvlapol}) slightly rotated the polarization position angle (Figure \ref{fig:angle}, top).
Removing residual polarization leakage did not significantly affect regions outside of the central synthesized beam area.
After removing the residual polarization leakage, the peak polarization intensity observed at 6.9 mm  is 400 $\mu$Jy\,beam$^{-1}$ ($\sim$17$\sigma$).
At 6.9 mm, the position angle of B-segments averaged over the synthesized beam at the peak of the polarized intensity map is 14$^{\circ}$ and 21$^{\circ}$ before and after removing the residual polarization leakage, respectively.
In order to make better comparison with the SMA data, the 6.9 mm map was made using the JVLA data in the limited $uv$ distance range of 10-455 $k\lambda$ and smoothed to the angular resolution of 0$\farcs$74. 
As shown in Figure \ref{fig:angle_smooth}, this image enhanced the significance of the detections of B-segments around IRAS4A2 (48$^{\circ}$$\pm$3.7$^{\circ}$), and the detections of B-segments around the lobes north (75$^{\circ}$$\pm$9.7$^{\circ}$), northeast (97$^{\circ}$$\pm$10$^{\circ}$), and southwest of IRAS4A1 (53$^{\circ}$$\pm$6.9$^{\circ}$; 4A1-N, 4A1-NE, 4A1-SW hereafter).

Around the IRAS4A region, the previous SMA $1\farcs6$$\times$$0\farcs99$ resolution observations at 0.88 mm reported the averaged B-segment position angle of 61$^{\circ}$ (Girart et al. 2006); and the previous Combined Array for Research in Millimeter-wave Astronomy (CARMA) $\sim$$2\farcs4$ resolution observations at 1.3 mm reported the averaged B-segment position angle of 56$^{\circ}$ (Hull et al. 2014).
The earlier Berkeley Illinois Maryland Association (BIMA) array $4\farcs4$$\times$$2\farcs8$ resolution observations at 1.3 mm detected the averaged B-segment position angle of 56$^{\circ}$$\pm$6$^{\circ}$ (Girart et al. 1999).
Close, but northern to the peak of 3 mm continuum emission, the earlier Owens Valley Radio Observatory (OVRO) $5\farcs1$$\times$$4\farcs3$ and $3\farcs7$$\times$$2\farcs8$ resolution observations at 3.1 mm reported the B-segment position angles of 13$^{\circ}$$\pm$6$^{\circ}$ and 7$^{\circ}$$\pm$6$^{\circ}$, respectively (Akeson \& Carlstrom 1997).

The observed B-segment position angles at 6.9 mm are only slightly deviated from those detected by the previous 1 mm and 0.88 mm observations (Figure \ref{fig:angle_smooth}).
We do not have an access to the OVRO data, and cannot know the detailed $uv$ sampling of the observations. 
It is difficult for us to know the exact reason of why the observed B-segment position angles by OVRO are largely offset from the results of other previous observations.
Presuming the offset is not due to calibration issues, we hypothesize that at $\sim$3 mm, the B-segment components observed by our JVLA observations at IRAS4A1 (more in Section \ref{chap_discussion}) already become significant.
At the central $\sim$1$''$ area, this B-segment component revealed by the long wavelengths observations shows a smaller (i.e. closer to 10$^{\circ}$-20$^{\circ}$) polarization position angle than that ($\sim$60$^\circ$) of the component which dominates the shorter wavelength observations.
Due to a finite synthesized beam area, the observed polarization position angles by OVRO may be a result of intensity-weighted sum among B-segments in the extended lobes (e.g. 4A1-N, 4A1-NE) and the B-segments within the central $\sim$1$''$ area around IRAS4A1.
In the case that the dominant contribution is from the central $\sim$1$''$ area, the averaged polarization position angle may become closer to 10$^{\circ}$-20$^{\circ}$).
% The observed polarization position angles by OVRO may be a result of intensity-weighted sum among the northwest-southeast B-segments in the north, and the northeast-south west B-segments contributed by the more extended envelope, due to a finite synthesized beam area.  }

At the observed spatial scales, a major difference between the results of the JVLA and the SMA polarization observations, is in spatial distribution of polarization intensity.
While the observed polarization intensity at 6.9 mm is centrally peaked (Figure \ref{fig:jvlaiquv}), the observed polarization intensity at 0.88 mm is peaked at and dominantly contributed by the two lobes northeast and southwest to IRAS4A1.
In fact, the observed polarization intensity appears to be avoiding the dense elongated structure connecting IRAS4A1 and IRAS4A2 (Frau et al. 2011; Lai et al. in prep.)
Our tentative interpretation will be given in Section \ref{chap_discussion}.
Both the JVLA and the SMA observations show the higher polarization percentage towards the outer regions. 
% interpretation: (1) flattened pseudo disk or accretion channel/filaments.  % Maybe also talk about 4B

\section{Discussion}
\label{chap_discussion}
Previous studies show that the continuum emission at 6.9 mm is dominated by dust thermal emission (Choi et al. 2011). 
The polarized continuum emission at submillimeter and millimeter bands, is likely attributed to emission of the elongated dust grains which are aligned perpendicular to the interstellar B-field. 
In this section, based on the interpretation given in Girart et al. (2006), Frau et al. (2011), and Kataoka et al. (2012) and references therein, we discuss potential reasons that may cause the discrepancy between the polarization observations at the 6.9 mm and the 0.88 mm bands. 
As mentioned in Section \ref{subsub:jvlapol}, our observational setups do not allow us to rule out the spatial variation of polarization leakage at off-center positions, although we do not expect this effect to be significant. 
The Stokes I peak of IRAS4B is $\sim$30$''$ separated from our pointing center, and was detected at S/N$\sim$100.
The non-detection of polarized emission (i.e. 3$\sigma$ limit) from IRAS4B implies that the polarization leakage both before and after the removal of residual polarization leakage  at its location is likely less than 3\%. 
If we assume that the off-center component of polarization leakage scales linearly with the angular separation from the pointing center, the estimated upper limit of off-center polarization leakage component will be $\lesssim$0.3\% within the IRAS4A region.

Our target source NGC1333 IRAS4A is a collapsing dense molecular core.
In this region, B-field is ordered on $\gtrsim$1000 AU scale, aligning in the northeast-southwest direction (Attard et al. 2009).
Within the central $\lesssim$1000 AU region surrounding IRAS4A1, the B-field is pinched in the geometrically flattened accretion flow which aligns perpendicular to the large-scale B-field.
In such a case, the observed polarization intensity distribution and polarization position angles at the 0.88 mm wavelength can present an hourglass geometry (Girart et al. 2006),  where the B-field is largely bent in the accretion flow converging from $\sim$1000 AU to $\sim$100 AU  regions; the B-field in the inner $\sim$100 AU region remains parallel with the large-scale B-field if the poloidal component is dominant on small scales, but is perpendicular to the large-scale B-field if the toroidal component is dominant.
% where B-field is largely bended in the accretion flow converging from $\sim$1000 AU to $\sim$100 AU region; the B-field in the inner $\sim$100 AU region remains parallel with the large-scale B-field, if the poloidal component is dominant.
The dominant emission of the 0.88 mm polarization intensity is from regions offset from the geometrically flattened accretion flow, which are northeast and southwest of IRAS4A1.
At 0.88 mm, the pinched, bended B-field in the flattened accretion flow becomes significant if generating polarization images with $uv$ distance range of $\gtrsim$50 $k\lambda$ (i.e. if filtering out the contribution from the extended envelope; Lai et al. in prep.)
Close to the plane of the dense accretion flow, radiation transfer models of Frau et al. (2011) and Kataoka et al. (2012) suggest that the polarized emission at 0.88 mm can be nicely canceled after being integrated in the line-of-sight, which leads to the low polarization percentage and low polarization intensity in this plane. 
The observed lower polarization percentages (Figure \ref{fig:diagrams}) in the higher density regions are consistent with this depolarization effect (Kataoka et al. 2012).  

For longer wavelength observations, either the effect of grain growth, or the effect of the much lower optical depth, can lead to a higher fractional contribution to the observed intensity from the dense central region (e.g. $<$200 AU), than from the $\sim$1000 AU scale envelope and flattened accretion flow.
These are consistent with the observed higher 6.9 mm to 0.88 mm flux ratios at regions which show brighter Stokes I emission.
% In Figure \ref{fig:diagrams}, we plot the observed polarization percentages at 6.9 mm and 0.88 mm versus the observed flux densities at those wavelength bands, and versus the 6.9 mm to 0.88 mm flux ratio.
As we show in Figure \ref{fig:diagrams}, for regions where we detect polarized emission in the two wavelengths, the 6.9 mm to 0.88 mm flux ratios and the Stokes I flux densities are correlated; the observed polarization percentages drop rapidly with both of these quantities.
The measured higher polarization percentage at longer wavelengths at the central region, and the observed centrally concentrated higher polarized intensity in the 6.9 mm observations, can be a consequence of the less serious canceling of polarized emission when integrating along the line of sight and within a finite synthesized beam.

We note that the recent JVLA $\sim$0$\farcs$2 resolution polarization observations at 8 mm and 10 mm have proposed that a toroidal B-field component may exist within the $\sim$$0\farcs5$ vicinity of IRAS4A1 (Cox et al. 2015).
We do not detect this toroidal B-field component.
We point out that the observed peak flux density of 12.7 mJy\,beam$^{-1}$ at 6.9 mm corresponds to a brightness temperature of $\sim$41 K, which is comparable to the averaged kinematic temperature at this region (Choi et al. 2010).
The 6.9 mm emission may be marginally optically thick within the central one beam area ($\sim$$0\farcs5$) at IRAS4A1; it becomes optically thin outside of this region and shows a  much lower brightness temperature (see Figure \ref{fig:jvlastokesI}).
At IRAS4A1, the 6.9 mm polarization observations may probe the B-field in a region nearer to the protostar than what was probed by the 0.88 mm observations.
Due to the high optical depth, the presented 6.9 mm, and the optical thicker 0.88 mm polarization observations may not yet probe the B-field inside the circumstellar disk in IRAS4A1  as seen by Cox et al. (2015).

We also point out that on the $\sim$2000 AU scale, the SMA polarization observations of CO J=3-2 have resolved the helical B-field from the wind-envelope interaction regions (Ching et al. 2016).
The observed orientations of B-segments around 4A1-N at 6.9 mm (Figure \ref{fig:angle_smooth}), although is deviated from the hourglass geometry, appear consistent with the helical B-field component detected in CO 3-2.

As a summary, we propose that the 0.88 mm polarized emission mainly traces the B-field in the extended envelope. 
The lower optical depth at 6.9 mm permits tracing the B-field in the inner part of the envelope, including the wind-envelope interaction regions, and the B-field closer in the $\sim$200 AU scale region.
The 6.9 mm emission is optically thick in the central $\lesssim$100 AU scale region, and cannot probe the B-field inside this region.
 
Our interpretation requires confirmation with additional JVLA observations (Section \ref{chap_summary}), or ALMA observations at 3 mm.
The quantitative modeling requires the details of the spatial distribution of dust and dust opacity spectral index (i.e. grain growth), the efficiency of grain alignment as a function of the localized physical environment (density, kinematics temperature, B-field strength) and grain size/morphology, which is beyond the scope of the present paper.
Tracing the distribution of large dust grains and the B-field configuration in the protostellar gas envelope may be required to understand ambipolar diffusion and magnetic braking self-consistently (Zhao et al. 2016)

The observed polarization percentages at 6.9 mm and 0.88 mm cannot yet be simply compared because of the different regions they probe. 
Our observations show that the polarization percentages at 0.88 mm and 6.9 mm are decreasing with the 6.9 mm to 0.88 mm Stokes I continuum flux ratio (Figure \ref{fig:diagrams}).
However, the significant, aforementioned depolarization effects on the observed spatial scales prohibits direct measurements of the effects of the rounding of growing dust grains (Hughes et al. 2009, 2013) on $>$100 AU scales. 
Close to the peak, the observed lower polarization percentage at 0.88 mm than 6.9 mm  (Figure \ref{fig:diagrams}) may be because of the more significant polarization canceling effect at 0.88 mm.
The high optical depth at 0.88 mm may further lead to additional (de-)polarization due to scattering (Kataoka et al. 2015; Yang et al. 2015).

% Since we do not have a map of dust temperature profile at the observed spatial scales, we are not yet able to distinguish the effects grain growing and optical depth.

\section{Conclusions and Remarks}
\label{chap_summary}
We have performed JVLA Q band (6.2-7.5 mm) polarization observations on the well studied Class 0 YSO NGC1333 IRAS4A. 
The Stokes I image resolves the well known YSOs IRAS4A1, IRAS4A2, and the two components in IRAS4B.
Towards the center of IRAS4A1, we detected (at $\sim$17$\sigma$) the polarization position angles slightly different from those observed from the existing SMA observations.
The detected polarization percentage by JVLA at 6.2-7.5 mm is $>$$\sim$3\% around the Stokes I continuum emission peak. 
Considering the worst case scenario that the flux is uniformly distributed in the synthesized beam of the presented Q band image, the 6.9 mm polarized intensity at this central region may be spatially resolved at 3$\sigma$ by 5 hours of the JVLA B array configuration observations ($\sim$$0\farcs15$).
The higher resolution polarization observations to probe the B-field in the inner regions will be preferably carried out in the longer wavelength bands, to avoid the high optical depth. 

The detected 6.9 mm polarized continuum emission is likely tracing predominantly the aligned dust grains.
Owing to the complicated B-field configurations in those lines of sight, and the higher optical depth for the 0.88 mm observations, it remains difficult for us to directly measure polarization percentage as a function of frequency, without being confused with the depolarization effects. 
The more sensitive observations to probe polarized 6.9 mm emission on extended regions where the B-field is ordered, can avoid the depolarization effects.

To our knowledge, this is the first successful JVLA observations of polarized dust emission at this frequency band.
This project also becomes a pathfinder for the future ALMA Band1 survey, which will have a similar sensitivity, but will benefit from the superior weather of the ALMA site.
Thus, our observations focus on the frontier area of observations of the polarized dust emission in the circumstellar environments, and contain an invaluable possibility of opening up a new window in synergy between JVLA, ALMA and SMA.

%============================================================
%============================================================

\acknowledgments
HBL thanks the supports from ASIAA.
We are extremely grateful to our referee Charles Hull, for the very useful comments.
HBL thanks Dr. Steve Myers for the information about polarization calibration sources, and thank Dr. Akimasa Kataoka, Francisca Kemper, Hiroyuki Hirashita, and Michihiro Takami, for useful suggestions. 
NH is supported by the MoST grant 104-2119-M-001-016.
J. M. G. acknowledges the support from MICINN (Spain) AYA2014-57369-C3-1-P grant and the MECD (Spain) PRX15/00435 travel grant.
C.C-G. acknowledges support by UNAM-DGAPA-PAPIIT grant number IA101214.
Y.H is currently supported by JPL/Caltech.

%============================================================
{\it Facilities:} \facility{JVLA, SMA} 

\end{document}